# Google Android: A State-of-the-Art Review of Security Mechanisms

Asaf Shabtai[1,3], Yuval Fledel[1,3], Uri Kanonov[1,3], Yuval Elovici[1,3] and Shlomi Dolev[2,3]

[1]Department of Information Systems Engineering, Ben-Gurion University, Israel.
[2]Department of Computer Science, Ben-Gurion University, Israel.
[3]Deutsche Telekom Laboratories at Ben-Gurion University, Israel.

---

**Abstract**. Google's Android is a comprehensive software framework for mobile communication devices (i.e., smartphones, PDAs). The Android framework includes an operating system, middleware and a set of key applications. The incorporation of integrated access services to the Internet on such mobile devices, however, increases their exposure to damages inflicted by various types of malware. This paper provides a comprehensive security assessment of the Android framework and the security mechanisms incorporated into it. A methodological qualitative risk analysis that we conducted identifies the high-risk threats to the framework and any potential danger to information or to the system resulting from vulnerabilities that have been uncovered and exploited. Our review of current academic and commercial solutions in the area of smartphone security yields a list of applied and recommended defense mechanisms for hardening mobile devices in general and the Android in particular. Lastly, we present five major (high-risk) threats to the Android framework and propose security solutions to mitigate them. We conclude by proposing a set of security mechanisms that should be explored and introduced into Android-powered devices.

**Keywords**: Mobile devices, Malware, Intrusion Detection System, Google, Android, Risk analysis, Security solutions for mobile devices.

---

## 1. INTRODUCTION

Designed as open, programmable, networked devices, smartphones are vulnerable to attacks. An infected smartphone can inflict severe damage to both users and the cellular service provider. Malware on a smartphone can: make the phone partially or fully unusable; cause unwanted billing; steal private information (possibly by phishing and social engineering methods); or infect every name in a user's phonebook [Piercy, 2005]. Attack vectors of malware propagating into smartphones include: Cellular networks, Bluetooth connection, Internet (via Wi-Fi, GPRS/EDGE or 3G network access), USB/ActiveSync/Docking and other Peripherals [Cheng, 2007].

The first smartphone virus, Cabir, was released in 2004 by the 29A virus writing group as a proof of concept of a self-replicating virus. Since then, several hundreds of smartphone viruses have emerged, many of which contain malicious codes and cause various levels of damage to smartphones. Smartphone malware evolve very quickly due to the experience virus writers have gained from the computer and Internet world. According to Gostev (2006), two years of smartphone virus evolution are equivalent to twenty years of work in computer viruses. Among the most prominent smartphone malware types are: the worm families -- Lasco/Cabir [Emm, 2005] and Commwarrior/Mabir [Schultz, 2006]; the Trojan viruses -- FlexiSpy, RedBrowser and Skulls; the Windows CE and CardTrap viruses [Leavitt, 2005]; and recently the iPhone ikee worm[1] and iPhone/Privacy.A hacking tool[2] that exploited vulnerabilities in jail-broken iPhone systems.

So far major pandemic outbreaks have been limited in scale due to the lack of a critical mass of victims. Nevertheless, since the smartphone market share is expected to increase significantly over the next few years, almost fivefold by 2013 [Frost and

---

[1] http://www.sophos.com/blogs/gc/g/2009/11/08/iphone-worm-discovered-wallpaper-rick-astley-photo/
[2] http://www.intego.com/news/hacker-tool-copies-personal-info-from-iphones.asp#note1

Sullivan, 2007], smartphones will provide a fertile ground for viruses. Another major factor attracting hackers is that smartphones are often carried for business purposes and are likely to have sensitive information. They also provide remote access to a company's most sensitive data, which can lead to data leakage if their phones are hacked into.

Among the most significant smartphone operating systems that have arisen recently is Google's Android framework. As a smartphone, the challenge for assuring Android's security is becoming similar to the personal computer [Muthukumaran, 2008]. Mobile OS makers are now very much concerned with the security challenges that PCs have been facing down through the years. The increasing number of attacks on mobile platforms (especially on smartphones) along with the increasing usage has led many security vendors and researchers to propose a variety of security solutions for mobile platforms. As a case in point, Symbian and Google have designed their operating systems to enable applications to run only in specialized sandboxes, minimizing the capability of malware to spread [Lawton, 2008]. A robust application signing and certification mechanism was integrated into Symbian's operating system and was proven highly effective in reducing malware attacks. This paper reviews and assesses the security mechanisms incorporated in Google's new Android framework. We provide a list of security mechanisms which can be incorporated to harden the Android. Based on this description we are able to make some recommendations on the efficacy and priorities of various security mechanisms.

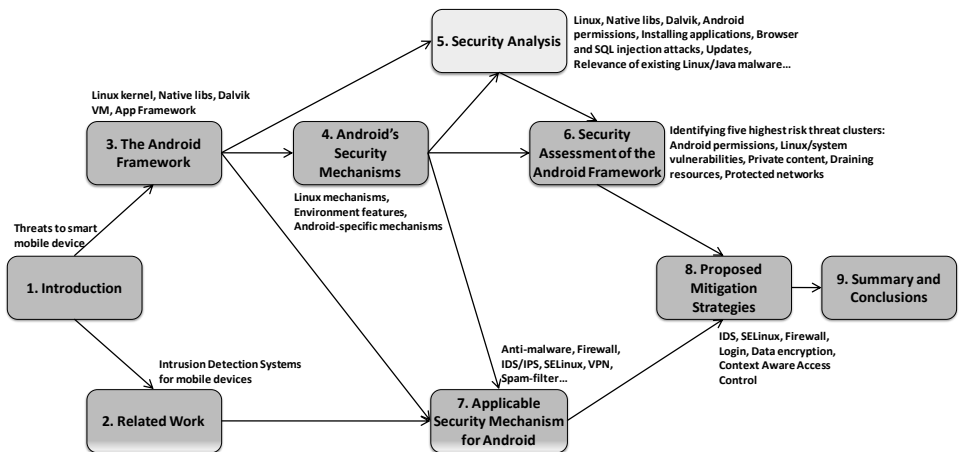

Fig. 1. Paper's structure

Figure 1 depicts the paper's structure. We start with description of related academic literature focusing on protection of mobile devices (section 2), description of the Android framework (section 3) and analysis of security mechanisms incorporated into it (section 4). In section 5 we describe the results of our security analysis and in section 7 we provide a list of security of applicable defense mechanisms for hardening mobile devices in general and the Android in particular. Section 6 describes the outcomes of a qualitative risk analysis that we conducted and presents five major (high-risk) threats to the Android framework. We conclude by proposing a set of security mechanisms that should be explored and introduced into Android-powered devices (section 8). Conclusions and future work are discussed in section 9.

## 2. RELATED WORK

Our overview of related academic literature indicates that most extant research on protection of mobile communication devices has focused on host-based intrusion detection systems. These systems, using anomaly detection or rule-based methods, extract and analyze (locally or by a remote server) a set of features indicating the state of the device. Several systems are reviewed in this section and summarized in Table I.

Table I: Academic research on protection of mobile devices.

| Paper | Approach | Detection method | Detects |
|---|---|---|---|
| Moreau, 1997 | HIDS | Anomaly detection using ANN | Fraudulent use of the operator services such as high rate calls. |
| Samfat, 1997 (IDAMN) | HIDS, NIDS | Anomaly detection; Rule-based detection | A user is being active in two different locations at the same time; traffic anomaly detection; and detecting anomalous behavior of individual mobile-phone users based on the telephone activity (such as call duration) and user's location in the network. |
| Yap, 2005 | HIDS | Signature-based detection | Proof of concept- detects unauthorized attempt to create SMS message. |
| Cheng, 2007 (SmartSiren) | HIDS, NIDS | Anomaly detection | Detects anomaly behavior of the device and outbreak of worm-related malware. |
| Schmidt et al., 2009 | HIDS | Anomaly detection | Monitor a smartphone running Symbian operating system and Windows Mobile in order to extract features for anomaly detection. These features are sent to a remote server for further analysis. |
| Bose, 2008 | HIDS | Signature-based detection. | Using Temporal Logic to detect malicious activity over time that matches a set of signatures represented as a sequence of events. |
| Kim, 2008 | HIDS | Signature-based detection | Detects, and analyzes previously unknown energy-depletion threats based on a collection of power signatures. |
| Buennemeyer, 2008 (B-SIPS) | HIDS, NIDS | Anomaly detection | Detects abnormal current changes and its correlation with network attack. |
| Nash, 2005 | HIDS | Statistical method (linear regression) | Detects processes that are likely to be battery-draining attacks. |
| Jacoby and Davis, 2006 (B-BID) | HIDS | Signature-based detection | Monitoring power consumption against "normal" power consumption range. Once an anomaly is detected, various system data is matched against known attack signatures. |
| Miettinen et al. (2006) | HIDS, NIDS | Event correlation | Combines both host-based and network-based data collection in order to be able to utilize the strengths of both detection approaches. |
| Hwang et al., 2009 | Authentication | Keystrokes dynamics | Collected 5 features to train and build a classifier capable of detecting impostor users. Utilized artificial rhythms and tempo cues to overcome problems resulting from short PIN length. |

Artificial Neural Networks (ANN) were used [Moreau, 1997] in order to detect anomalous behavior indicating a fraudulent use of the operator services (e.g. registration with a false identity and using the phone to high tariff destinations) based on 16 features representing mean and standard deviation of the total duration and number of long/short term national and international calls.

The *Intrusion Detection Architecture for Mobile Networks* (IDAMN) system [Samfat, 1997] uses both rule-based and anomaly detection methods. IDAMN offers three levels of detection: location-based detection (a user active in two different locations at the same time); traffic anomaly detection (an area having normally low network activity experiencing high network activity); and detecting anomalous behavior of individual mobile-phone users. In order to detect anomalous behavior, a profile is generated by monitoring the user's telephone activity (e.g., call duration, inactivity time between two calls, number of handovers performed, total number of calls in a session and total duration of calls in a session). In addition, the user's location in the network (roaming) is monitored by generating a state machine with the probability of moving from one location (cell) to another.

Yap et al. (2005) employ a behavior checker solution that can detect malicious activities in the system. They present a proof-of-concept scenario using a Nokia Mobile phone running a Symbian OS. In the demonstration, a behavioral detector detects a simulated Trojan attempting to use the message server component without authorization to create an SMS message.

Cheng et al. (2007) present *SmartSiren*, a collaborative proxy-based virus detection system. Single-device and system-wide abnormal behaviors are detected by the joint analysis of communication activity of monitored smartphones. The SmartSiren architecture consists of a back-end proxy that interacts with lightweight agents on the protected devices. The agents merely collect information and relay it to the proxy that performs the analysis and sends out the alerts.

Schmidt et al. (2008) monitored a smartphone running a Symbian OS by extracting features that describe the state of the device and which can be used for anomaly detection. These features were collected by a Symbian monitoring client and forwarded to a Remote Anomaly Detection System (RADS). The gathered data was analyzed in order to distinguish between normal and abnormal behavior. The results indicated that most of the top ten applications preferred by mobile phone users affected the monitored features in different ways.

An interesting behavioral detection framework is proposed in [Bose, 2008] to detect mobile worms, viruses and Trojan horses. The method employs a temporal logic approach to detect malicious activity over time. A malware behavior is represented by describing the temporal ordering of an application's actions which may reveal malicious intent even when each action alone may appear harmless. A database of malicious behavior signatures was generated by studying more than 25 distinct families of Symbian OS malware. Next, a two-stage mapping technique constructs these signatures in run-time from the monitored system events and API calls in the Symbian OS. The system differentiates the malicious behavior of malware from the behavior of benign applications by training a classifier based on Support Vector Machines (SVM).

Special effort has been devoted to Intrusion Detection Systems (IDS) that analyze generic battery power consumption patterns to block Distributed Denial of Service (DDoS) attacks or to detect malicious activity via power depletion [Racic, 2006]. Martin et al. (2004) propose three main methods for an attacker to drain the battery: (1) Service request attacks, where repeated requests for services are made to the victim, typically over a network. The target must expend energy handling the request even if the service is not provided; (2) benign power attacks, where a victim is induced to repeatedly execute a valid but energy-consuming task; and (3) highly intrusive attacks, where the attacker modifies an executable to make it consume more energy than actually required.

Kim (2008) presents a power-aware, malware-detection framework that monitors, detects, and analyzes previously unknown energy-depletion threats. The framework collects power consumption samples and generates power signatures. These signatures are used for classifying mobile malware by measuring the similarity between power signatures using the χ2-distance measure.

The *Battery-Sensing Intrusion Protection System* (B-SIPS) [Buennemeyer et al., 2008] for mobile computers alerts when abnormal current changes are detected. The B-SIPS correlates host-based anomaly detection with the Snort IDS which provides signature-based detection of attack.

Nash et al. (2005) presented a design for an intrusion detection system that estimated power consumption according to a linear regression model, based on parameters such as CPU load and disk accesses, to determine the amount of energy used on a per process basis and to identify processes that could potentially exhaust batteries.

Jacoby and Davis (2006) presented a host *Battery-Based Intrusion Detection System* (B-BID). The basic idea is that monitoring the device's electrical current and evaluation the correlation with known signatures and patterns can facilitate attack detection.

Miettinen et al. (2006) claimed that host-based approaches are required, since network-based monitoring alone is not sufficient to encounter the future threats. They adopt a hybrid network/host-based approach. A correlation engine on the back-end server filters the received alarms according to correlation rules in its knowledge base and forwards the correlation results to a security monitoring GUI to be analyzed by security administrators.

Hwang et al. (2009) evaluated the effectiveness of Keystroke Dynamics-based Authentication (KDA) on mobile devices. Their empirical evaluation focused on short PIN numbers (four digits) and the proposed method yielded a 4% misclassification rate.

## 3. THE ANDROID FRAMEWORK

Android[3] is an application execution environment for mobile devices that includes an operating system, application framework and core applications. The applications are written in Java based on the APIs provided by the Android Software Development Kit (SDK). The foundation of the Android software stack is the *Linux Kernel*. Android uses Linux for its device drivers, memory management, process management, and networking. Usually, an application developer will not be directly programming to this layer. The next level up contains the *Android Native Libraries*. These libraries are written in C/C++ and are used by various system components in the upper layers. Incorporating these libraries in Android applications is achieved through Java interfaces. This layer contains a customized C-library, an SQL database engine, 2D and 3D graphic libraries, a native web browser engine (WebKit) and media codecs (e.g., MPEG-4 and MP3). Next is the *Android Runtime*, consisting of the Dalvik Virtual Machine and the core libraries. Dalvik runs .dex (Dalvik Executable) files which are more compact and memory-efficient than Java class files. This is an important consideration for battery-powered devices with limited memory. The core libraries are written in Java and provide a substantial subset of the Java 5 SE packages (e.g., standard collections, I/O, networking, utilities) as well as some Android-specific libraries which are needed for accessing the capabilities that the hardware, operating system and native libraries offer. The *Application Framework* layer, written fully in Java, includes Google-supplied tools as well as proprietary extensions or services. An important component of the framework is the Activity Manager, which manages the lifecycle of applications. The top layer is the *Applications* layer for implementing such applications as a phone, web-browser, email client and more. Each application in Android is packaged in an .apk (Android package) archive for installation. The .apk is similar to a standard Java jar file in that it holds all code and non-code resources (e.g., images, manifest) for the application. The Android package is a collection of components. Components in one apk are isolated from components in another apk and can only communicate with each other and share data through means provided by the system. Each apk is associated with a primary process in which all of the application's components (i.e., Activities, Services, Broadcast Receivers and Content Providers) are executed. These application's components, along with capabilities, permissions and requirements should be listed in the AndroidManifest.xml file. Detailed description of the application's components and related security properties is provided in [Enck, 2009b].

---

[3] http://www.android.com/

## 4. ANDROID'S SECURITY MECHANISMS

Android is a multi-process system, where each application (and parts of the system) runs in its own process. For the most part, security between applications and the system is enforced at the process level through standard Linux facilities, such as user and group IDs assigned to applications. In addition, access-control is provided through a permission mechanism that enforces restrictions on the specific operations that a particular application can perform. Several security aspects and mechanisms are bundled by Google in the Android framework. Enck (2009b) and Burns (2008) present the main components of an Android application and how the Android-specific mechanism can be used correctly to protect an Android application. We cluster them into three general groups: Linux mechanisms, environment features and Android-specific mechanisms (Table II).

### 4.1 POSIX (Portable Operating System Interface) Users

Each Android package (.apk) file installed on an Android-powered device is given its own unique Linux (POSIX) user ID. This user ID is assigned when the application is installed on the device. Consequently, the code of two different packages cannot run in the same process. In a way, this creates a sandbox that prevents one application from touching other applications (or, vice versa, other applications from touching it). In order for two applications to share the same permissions set and possibly run in the same process, they must share the same user ID, which is only possible through the use of the sharedUserID feature. To share the same user ID, two applications must explicitly declare the usage of the same sharedUserID and both must bear the same digital signature (see section 4.8).

Seeing as an application's code always runs in its own process, regardless of how it was invoked (by that application directly or by a form of communication from another application), the permissions the code is run with are that of the owning application (both regarding file access and application-level permissions). For example in a scenario where the Contacts application open's the SMS application's editor, the editor would both be able to send an SMS and access any files owned by the SMS application whereas the contacts application that triggered the starting of the editor could do neither.

### 4.2 File Access

Files in Android (both application- and system-files) are subjected to the Linux permission mechanism. Each file is associated with the owner-user and group IDs and three tuples of Read Write and eXecute (rwx) permissions. The first tuple is enforced on the owner; the second on users that belong to the group; and the third is for the rest of the users. Those permissions bits are enforced by the kernel. Generally, system files in Android are owned by either the "system" or "root" user, and application files are owned by an application-specific user. The access permissions for files are derived from the

Linux's user mechanism and reflect the level of access permitted to other users (i.e., other applications). In the same manner, due the difference in user IDs, system files are protected from ordinary applications. Files created by an application will be assigned that application's user ID, and will not be accessible to other applications (unless they share the same user ID through the sharedUserID feature or files are set as globally readable/writeable). Linux provides many system functionalities as semi-files. This mechanism effectively allows setting permissions on files, directories, drivers, terminals, hardware sensors, power state changes, audio, direct input readings, shared memory, and access to background daemons.

One aspect that reinforces these various security measures is that the system image is mounted as read-only. All important executables and configuration files are located on either the ramdisk (which is read-only too, but also reinitialized from a known state on every boot) or the system image. Therefore, an attacker that gains the ability to write to files everywhere on the file system is still denied the possibility of replacing critical files. However, an attacker can increase the attack's complexity and overcome this limitation by remounting the system image. This, however, requires root access.

Two other interesting partitions on the Android file system are the data partition and the SD card. The data partition is where all user data and applications are stored. Since this partition is distinguished from the system partition, it effectively sets a quota on the amount of user data that can be input or loaded into the device. This effective quota prevents the system partition from damage in case the user (or a malicious attacker) installs too many applications or creates too many files. When the Android device is started in "safe mode", files from the data partition are not loaded, making it possible to recover from such attacks. The SD card is an external storage device, and therefore can be manipulated off-line, and out of the control of the device. The only file system that is supported on the T-Mobile G1 phone for such a card is the FAT file system. The FAT file system does not support UNIX-like permissions and file ownership, both of which can only be set through mount options.

### 4.3 Memory Management Unit (MMU)

A prerequisite of many modern operating systems and Linux in particular, is the Memory Management Unit (MMU), a hardware component which facilitates separation of processes to different address spaces (virtual memory). Various operating systems employ the MMU in such a way that one process is unable to read the memory pages of another process (information disclosure), or to corrupt its memory. The likelihood of privilege escalation is reduced since a process is unable make its own code run in a privileged mode by means of overwriting the private OS memory.

### 4.4 Type Safety

Type safety is a property of programming languages. It forces variable contents to adhere to a specific format and therefore prevents erroneous or undesirable usage. Partial or missing type safety and/or boundary checks can lead to memory corruption and buffer overflow attacks, which are the means to arbitrary code execution. As noted, Android employs Java, which is a strongly typed programming language. Programs written in this environment are less susceptible to arbitrary code execution. This is in contrast to such languages as C, which allows casting without type checking, and performs no boundary checks unless they are specifically written by the programmer. Android allows programs to have native components written in C, at the risk of potentially reduced security. Binder, the Android-specific Inter-Process Communication (IPC) mechanism, is also type safe. Types of data elements that are passed by Binder are defined by the developer on compile time in Android Interface Definition Language (AIDL). This ensures that types are preserved across process boundaries.

### 4.5 Mobile Carrier Security Features

A basic set of attributes of telephony systems, and utilities in general, come from the need to identify the user, monitor usage and charge the client accordingly. A more general term is *AAA*, which stands for Authentication, Authorization and Accounting. As a smartphone, Android borrows these classical security features from cellular phone design. *Authentication* is usually done by a SIM card and associated protocols. The SIM card contains a secret shared only by the card and the operator.

Table II. Security mechanisms incorporated in the Android.

| Mechanism | Description | Security issue |
|---|---|---|
| **Linux mechanisms** | | |
| POSIX users | Each application is associated with a different UID. | Prevents one application from disturbing another. |
| File access | Application's directory is only available to the owned application. | Prevents one application from accessing files of another. |
| **Environment features** | | |
| Memory Management Unit | Each process is running in its own address space. | Privilege escalation, information disclosure, DoS. |
| Type safety | Enforcing variable content to adhere to a specific format, both in compiling-time and runtime. | Prevents buffer-overflows, and stack smashing. |
| Mobile Carrier Security Features | Using SIM card to authenticate and authorize user identity. | Phone call theft. |
| **Android-specific mechanisms** | | |
| Application permissions | Each application declares which permission it requires at install time. | Limiting application abilities to prevent malicious behavior. |
| Component Encapsulation | Each component in an application (e.g., Activity, or Service) has a visibility level that regulates access to it from other applications (e.g., binding to a service) | Prevents one application from disturbing another, accessing private components or APIs |
| Signing applications | Application apk files are signed by the developer and verified by the package manager. | Matching and verifying that two applications are from the same source. |
| Dalvik VM | Each application runs in its own virtual machine. | Prevents buffer-overflows, remote code execution, and stack smashing. |

### 4.6 Application Permissions

The heart of the application level security in Android is the permission system which enforces restrictions on specific operations that an application can perform. The Package Manager is in charge of granting permissions to applications at installation and the application framework is in charge of enforcing system permissions on runtime.
There are about a 100 built-in permissions in Android which control operations including: dialing the phone (CALL_PHONE), taking pictures (CAMERA), using the Internet (INTERNET), listening to key strokes (READ_INPUT_STATE) or writing an SMS (WRITE_SMS). Any Android application can declare additional permissions. In order to obtain permission, an application must explicitly request it in its manifest (the application's "contract" with Android). Permissions have associated protection levels: (1) Normal (permissions that are not especially dangerous to possess); (2) Dangerous (permissions that are more dangerous than normal, or not normally needed by applications; such permissions may be granted to an application with the user's explicit confirmation); (3) Signature (permissions that can only be granted to other packages that are signed with the same signature as the one declaring the permission); and (4) SignatureOrSystem (a signature permission that is also granted to packages installed in the Android system image).

The assignment of protection level is left to the developer's will, intuition or common sense. However, the following guidelines are useful in the process. "Normal" permissions should imply minor risk and serve only as a "heads-up" for the user that the application is requesting access to such functionality. "Dangerous" permissions on the other hand, should be used for operations implying a more substantial risk. A careful balance needs to be maintained between the two categories to avoid misclassifying dangerous operations as low-risk which may result in the unwise granting of access to the functionality. The other extreme (stating everything is "dangerous") is just as risky, and may cause the user to ignore the importance of the protection level due to lack of contrast. Contrary to the "Normal" and "Dangerous" levels, "Signature" permissions are intended only for operations intended to be used by applications from the same developer ("Same-origin" policy). Lastly the "SignatureOrSystem" level is discouraged as a whole.

At installation, permissions requested by the application are granted to it based on checks against the signatures of the applications declaring these permissions and interaction with the user. After the application has been installed and its permissions have been granted (perhaps partially) it can no longer request any further permissions and no permissions, once denied, can be granted since there is no further interaction with the user after the installation. An operation which has not been granted permission will fail at run-time. Thus, when installing an application, the user has but two choices - trust the developer or not. If he decides to trust the developer (i.e., trust the application) all permissions will be granted. Otherwise, the only option is not installing the application. In such a case the factor with the most influence on the user's choice is his craving for the application requesting the permissions.

In addition to guarding protected framework APIs, the permission mechanism can and should be used to secure the various components in an application which are the Activity, Service, Content Provider and Broadcast Receivers. This effect is achieved primarily by associating permissions with the relevant component in its declaration in the manifest and having Android automatically enforce for the existence of the permissions in the relevant scenarios. An *Activity*, which is the UI building block of an Android application (any screen a user sees on the device and interacts with), can specify a set of permissions that will be required for any applications wishing to start it. In a similar manner a *Service*, which is the background building block of Android applications, can control which applications will be allowed to start/stop it or to bind to it. Whereas such permissions provide a crude restrictions, a more fine-grained control of any API exposed through binding can be obtained through run-time checks of the permissions granted to the bound application.. *Content Provider* provides the means to store and share data. It may define permissions to regulate who is allowed to read or write the information associated with the Content Provider. Since the read and write permissions are defined separately and are not interconnected, the Content Provider has fine-grained control over access. *Intent Broadcasts* (what an application wants to be done) and *Broadcast Receivers* (executed by the system in reaction to Broadcasted Intents) can both be associated with a set of permissions. For the Broadcast Receiver this feature makes it possible to control which components are allowed to broadcast the Intents it is configured to receive. For the component that is broadcasting the Intent, this associative feature provides the ability to control which Broadcast Receivers will receive it. The implication is that any attempt to spy on the Intents being broadcasted is severely crippled and efforts to spoof broadcasts by unauthorized components will fail.

### 4.7 Component Encapsulation

By providing an application with the ability to encapsulate its components (i.e., Activity, Service, Content Provider and Broadcast Receiver) within the application content, Android disallows any access to them from other applications (assuming that they bear a different user-ID). This is primarily done by defining the "exported" feature of the component. If the "exported" feature is set to false, the component at hand can only be accessed by the owning application and any other application sharing its user-ID through the shared user-ID feature. If set to true, it can be invoked by external applications. However, the invoking applications may still be controlled through the permission mechanism as described in the above section. It is prudent to always set the "exported" feature manually and not to count on the default behavior of the system since it may not coincide with the expected behavior. Naturally, whether a component is encapsulated or not, has no effect on its ability to access other components, which is limited only by their encapsulation and access permissions.

### 4.8 Signing Applications

Each application in Android is packaged in an .apk archive for installation. The .apk archive is similar to a Java standard jar file in that it holds all the code (.dex files) for the application. In addition it also contains all the application's non-code resources such as images. The Android system requires that all installed applications must be digitally signed (code and non-code resources).
The signed apk is valid as long as its certificate is valid and the enclosed public key successfully verifies the signature. Signing applications in Android is used to verify that two or more applications are from the same owner ("same-origin" verification). This feature is used by the sharedUserId mechanism and by the permission mechanism to verify Signature and SignatureOrSystem protection level permissions.

## 5. SECURITY ANALYSIS

In this section we describe findings from our comprehensive assessment of various security aspects of the Android framework which were validated on a T-Mobile G1 device. Among the elements we analyzed are: the codes of various Android components; the application permission-granting mechanisms and application installation process; and the applicability of existing Linux and Java malware.

### 5.1 Analysis of the Android Framework's Cornerstone Layers

This subsection describes the outcome of our analysis of Android's lower layers. We adopted a security-oriented, code-review approach to identify potential vulnerabilities. It should be noted that Android includes a substantial number of lines of code and we did not attempt to review and look for all vulnerabilities/bugs in the code. Rather we focused on special locations that might be problematic such as interfaces, structures etc. The code that we reviewed is based on the USA variant of the T-Mobile G1 device, version RC30.

*5.1.1 Linux Kernel*

The Linux kernel is not famous for its security. In fact, in 2007, 83 CVE (Common Vulnerabilities and Exposures) entries were logged on it, a little over one entry a week. In 2008, 73 CVE entries were logged. The drivers and vendor-specific additions serve as two locations that are particularly "hospitable" to bugs. Drivers and subsystems run with the highest privileges. A vulnerability that is detected and exploited by the hacker will usually achieve his/her goal, i.e. "root access". From the results of our analysis, we conclude that proper security countermeasures should be added to this layer. The following paragraphs list several decisions taken while implementing Android on the G1 device and discuss their security implications.

**Submitting code to mainline**: Code that is integrated into the mainline passes several phases of checks and validations. When code is reviewed, bugs are more than likely to surface; some may have security implications. In this regard, Android has diverged from the mainline kernel by adding (or extensively modifying) vendor-specific drivers and modules that were not reviewed. Some are not likely to be merged without serious redesign. Additionally, modifications implemented by Google on top of the Linux kernel were published in a deployed platform before attempting to submit the changes upstream. Examples of such modifications include the Android shared memory driver (Ashmem) that allows applications to share and manage memory at the kernel level and an Inter-Process Communication mechanism (Binder).

**Development phase**: Developers are required to verify that development-stage backdoors and logging components are removed before deployment in the production environment. But in two cases, this best-practice was never properly verified. In the first example, all versions up to and including RC29 had a debugging-phase code which sent to a background root shell every key that was pressed on the keyboard. This back-door on the G1 device enabled users to gain full control of the system (i.e., root user). A second example is the kernel logging facility (dmesg) that developers used but which may lead to a Denial of Service (DoS) by too-verbose drivers that do not rate-limit their logging messages. The Binder is a notable example of an overly-verbose driver. This issue was handled recent Android firmware and many logging messages were removed.

**Modification and extension of existing functionality**: Some of Google's modifications extend existing functionality. In some cases a better choice would have been to integrate these modifications into other components. Examples of such components are the *Ashmem* which extends Shmem and the *Lowmemorykiller* which extends the Out-Of-Memory (OOM) killer. Shmem is a standard POSIX feature which allows multiple processes to share memory. Ashmem, however, is an Android wrapper over Shmem which uses the ability of the kernel to release shared-memory allocations when the system is tight on memory. The standard OOM-killer, under tight memory conditions, first asks kernel subsystems to voluntarily free memory. If no memory is freed, the OOM-killer resorts to killing a process. The decision as to which process to kill is based on multiple attributes such as consumed memory, CPU time, permissions etc. Android's lowmemorykiller is more aggressive since it kills a process without regard to whether memory has been freed voluntarily or not.

**Hardcoded POSIX users and groups**: Android modifications include hardcode user ids (uids) and group ids (gids) in the kernel code. These modifications contradict the basic design decisions in Linux. However, Android is not a general Linux platform and using hardcoded group IDs to manage security configuration without extensive additional security infrastructure is a legitimate solution. In fact, it increases security since the system services are not required to run with root privileges. Two examples of such modifications are the "paranoid network" set of changes that limits network access based on gids, and the Binder which accepts the first process that uses it as its master, but only if it has the "system" group.

**Kernel Configuration**: The Linux kernel is highly configurable. On the G1 device, many common Linux options are disabled in order to reduce memory consumption. Less code means a reduction in vulnerability. On the other hand, it also means omitting "security enabler" modules. A few examples that have security implications are: disabling the auditing support and BSD task accounting (less input is available to the host-based intrusion detection system); disabling SYN cookie support (if enabled, it can reduce the chance of SYN flood attack); enabling PPTP, L2TP and IPSec-based VPN connections (on Android release 1.6; i,e., Donut); enabling the CFS scheduler group scheduling (no application is able to monopolize the CPU); disabling security modules (such as SELinux) and enabling NetFilter (allowing the implementation of a host firewall).
Enabling these modules in the kernel configuration requires a trivial amount of effort (and also consumes minor memory space); although, providing the means for using these modules is not trivial. This requires additional user-space components that will provide the API for such modules in order to avoid the need for root access. As an example, the SELinux module requires code and script modifications [Shabtai, 2009c].

*5.1.2 System Libraries*

Android makes use of many native libraries. These libraries are intended to be used by native processes, other native libraries or by Dalvik through Java Native Interface (JNI). JNI is a method for calling native methods from Java in the context of the same process. JNI is normally used for: (1) providing low-level functionalities (e.g., string operations, socket access, file handling, thread creation, Inter-Process Communication); (2) implementing computationally intensive calculations (e.g., handling multimedia and 3D graphics); (3) hiding code and licensing issues; and (4) leveraging existing libraries.
The native libraries are written in C/C++, which is not type safe. Thus, native libraries have a higher chance of bugs than Java code. Since JNI loads native libraries in the memory space of a Dalvik process, bugs in the native library may crash the Dalvik process, corrupt its memory or cause arbitrary code execution. For that reason, system libraries are a target when searching for security vulnerabilities.
At both the development stage and following the release of the first device, Android's native libraries had already been exploited. The main sources of these exploited libraries were outdated vulnerable versions of ported libraries, such as the SQLite (a lightweight relational database engine), Webkit (Web browser engine) and the new implementation of native libraries. Writing libraries from scratch means that bugs were not flushed out. Perhaps even more reprehensible from a security perspective is that Google published Android's native libraries code just as the first device hit the market.

*5.1.3 Dalvik Runtime*

Dalvik is a Java Virtual Machine (VM) based on Apache Harmony which was extensively modified and adapted for environments with low memory. Dalvik is a complex component with approximately 73k lines of C/C++ code, 8k of assembly and 390k lines of Java (not including auxiliary utilities or unit tests). Since much of it is new

(including the .dex file format the VM runs), it needs to be screened for security problems. Dalvik provides the possibility of executing native code through JNI without requesting permission for it. Employing native code, however, should be used wisely since it removes the layer of defense provided by the VM.

Securing Dalvik is crucial since vulnerability in the VM affects all applications. A potential weak spot is Dalvik VM .dex file loading code which is required to deal with .dex files from unreliable sources.

By inspection of the Dalvik code we conclude that sanity checking is implemented in the initial loading code. Additionally, many "assert" statements are visible, which log a message when a sanity check fails. However, "assert" statements are useful for development and QA, but not for stopping maliciously malformed files. Furthermore, we also found pointer arithmetic that was not sanity checked (e.g., the pointers to the strings table, types table, class tables). As a result, we were able create a malformed .dex file that during installation caused the Package Installer to crash resulting in a phantom application that cannot be uninstalled because the installer claimed it is already installed, nor can another package with the same package name be installed if it has a different signature.

The verification process of .dex files that is performed during the installation of the application is also applied whenever the .dex files are loaded to memory. Another lightweight verification is performed at run-time. Dalvik's bytecode verifier does not enforce Java's type safety. As a result an attacker may be able to compile his own type-unsafe code to .dex bytecode and to run such bytecode. This may cause to the application to crash, or run "arbitrary" code. This is not a concern, seeing as: 1) the attacker will crash his own application/malware; and 2) both effects can be achieved by simpler means not involving the engineering of a malformed .dex file. Thus, malformed .dex files would not give the attacker any new capabilities in this case.

A major design decision in regard to Dalvik was that it would only start once, and then get cloned. This initial "Zygote" process gets commands from the system server and acts accordingly. We found that an explicit check of the origin of commands, and commands from unreliable sources will raise an Exception.

## 5.2 Application-Level Permissions

Since Android is an open framework, device-holders are in the position of enjoying the wide variety of applications and services that developers are providing and will provide in the future. The downside is that it will be hard to inspect and block certain applications that will be provided by unreliable sources. It is very likely that a typical Android user will download unsafe software and simply click through various warning messages.

The application-level permission mechanism is responsible for securing various APIs provided by the system and other applications. Whereas many of the core permissions are reserved for Google applications (due to their Signature or SignatureOrSystem protection level), a large variety of Normal and Dangerous protection level permissions are still available for non-Google applications. As a result abuse of such permissions is inevitable.

An Internet access with the ability to read various contents stored on the device (e.g., SMS messages, call logs, contacts; GPS tracking; audio recording; and camera features) can be used to acquire confidential information or to spy on the user without his/her knowledge. Other malware-induced possibilities include: Denial of Service (DoS) attacks, such as denying the ability to place non-emergency number phone calls; draining the battery; termination of applications; and the abuse of chargeable services (e.g., phone calls, SMS/MMS messages, and chargeable network traffic).

Another source of difficulties arises from the shared user-ID feature. When an application declaring a shared user-ID is installed, all of the granted permissions are ascribed to the shared user-ID. At runtime, each of the applications sharing that user-ID will be granted by a combined set of permissions. A simple attack scenario based on exploiting the user-ID feature would probably take place as follows. The user installs two applications sharing a user-ID. The first requests access to the Internet while the second wishes to access the contact list. As soon as both applications are installed, each application is capable of both reading the contacts and sending them through the Internet. The user, however, is unaware of the collaboration between the completely unrelated applications.

A generalization of the above scenario may be seen as two seemingly unrelated applications that collaborate to leak information from the device using a shared medium to transfer the information between the two. In the shared user ID scenario there was no need for a shared medium seeing as both applications were "as one", however such a medium may be a Content Provider, an Intent, a Service with an exposed API or even a plain socket.

### 5.3 Installing Applications

Android applications are distributed as .apk (Android Package) files and the actual installation process can be perceived as the deployment of the .apk on the device. The .apk can be regarded as an archive containing the applications code (a .dex file) and non-code resources.

The Package Manager (the service responsible for the installation process) validates the correctness of the apk. Validation includes but is not limited to: verification of the digital signature; confirmation of legitimacy of shared user-ID or permission requests; and the validation/verification of the included .dex file. Due to the lack of a certificate authority (CA), it is not possible to verify the identity of the developer and consequently the integrity of the .apk. As a result, unless the application uses signed-related features (permissions or shared user-ID) any tampering with the .apk will not be detected. The package installation API is guarded by the INSTALL_PACKAGES permission which is of Signature protection level and defined in the core Android package. Thus, only Google-signed applications may serve as application-level wrappers for the Package Manager; i.e. malicious applications cannot install applications on their own. The "Package Installer", a legitimate application-level wrapper for the Package Manager is included as one of the core applications that are supplied with the Android-operated device.

There are three main methods for installing apk files on an Android device. They differ in how they obtain the .apk files and whether they interact with the Package Installer application or the Package Manager directly. The Android Debug Bridge (a command-line tool that is supplied along with the SDK) is intended only for developers. The installation command is issued from a PC which is connected to an Android device using a USB cable. The actual installation is done directly by the Package Manager without any user interaction and resulting in automatic granting of Normal and Dangerous permissions. The lack of user interaction makes the process a highly risky way of installing applications and has a high impact on the device's security. Therefore, this method should not be used on applications with which the developer is not familiar.

The two remaining installation options are intended for the general public, the primary and officially sanctioned one is installing via Android Market (a Google-proprietary application which allows browsing and downloading of applications that were published by different developers). Since the Market application is signed by Google, it interacts directly with the Package Manager. The last installation method is based on installing applications from the SD card. There are a number of free, 3rd party applications that enable the user to explore the SD card in search of .apk files and to initiate the installation process which is carried out by the Package Installer (e.g., ApkInstaller and AppsInstaller). Unlike applications obtained from the Android Market, the installation of applications from unknown/unreliable sources requires a confirmation from the user that he/she is aware of the risks involved in installing applications from unknown sources. Thus, by default, the installation of applications from unknown sources is prohibited. It should be noted that this restriction is enforced by the Package Installer and not by the Package Manager. Therefore it doesn't apply to methods that bypass the Package Installer and uses the Package Manager directly.

When installing an existing application, the installation will be allowed only when the signatures of the existing application and the new application match. The signature-matching safeguards against malicious applications that attempt to gain access to private data through substitution of the original applications.

A major design flaw in both the Package Manager and Package Installer is that the user does not have the ability to only partially grant the requested permissions. If the user chooses to grant all permissions, only non-matching Signature (or SignatureOrSystem) permissions will not be granted.

### 5.4 Web-Browser

Web-browsing exposes Android users to common attacks such as: Cross-Site Scripting (XSS); URL encoding attacks; social engineering; and malicious scripts. WebKit, Android's open-source Web engine, has a history of vulnerabilities. Some recent attacks on the Web browser include a buffer overflow in an outdated native library, and an explicit XSS vulnerability. Both attacks enabled the attacker to run any malicious code on the device with all the abilities and privileges assigned to the Web browser application. Since, like any Android application, the browser runs with its own POSIX user-id, an attack is limited to the browser, leaving other phone functions (e.g., dialing or messaging) unharmed. The browser is also limited by the application-level permissions it has been granted at installation (Internet access, ability to acquire wake locks, location-based APIs and network-related information retrieval). Nevertheless, in a successful attack, the attacker could gain information stored by the browser such as cookies, passwords, favorites and form-field values. Having access to all of the browser's private data, the attacker could corrupt it in order to prevent correct operation in the future.

The browser provides several security-related configurable options. These include remembering form data and passwords; accepting cookies; displaying security warnings; loading images; enabling JavaScript; blocking pop-ups; and setting (or disabling) the homepage. In addition, the browser prevents any downloads of files whose MIME type cannot be handled by the device, e.g. ZIP or RAR. However, since the MIME type is set by the server this filtering mechanism can be bypassed by having the server report a false MIME type in the HTTP response. A proof-of-concept scenario was tested on the G1 by downloading an actual ZIP file under the disguise of an apk.

### 5.5 SQL Injection

SQLite, the most common persistent storage provider on the Android platform, is used by the vast majority of the system and user applications. Consequently, we have reviewed the SQLite module as well as related code in the Android framework in order to understand the level of Android's exposure to SQL injection attacks. The API of the Android SDK works automatically or manually with SQLite databases with respect to embedding user-inputted data into queries. Manual embedding is achieved by appending the parameters to an SQL query string. This method is susceptible to SQL injection. An alternative is to automatically bind parameter values in a safe manner. It is possible to embed place holders in queries and separately pass on their values at execution time. The SQL injection issue is handled by binding the values to their respective place holders on a much lower level (the SQLite native library) and by embedding them not as strings but according to their binary values. This method renders inert any attempt to use escape characters in the input since they are treated like any other character. We also inspected the use of the SQL query and update APIs in the Android code-base (excluding 3rd party applications). Our finding shows that the API encourages writing clients in the non-vulnerable approach (i.e. automatic binding) and the vast majority of the code use the safe API. In conclusion, the responsibility of mitigating SQL injection is left in the hands of the developer and only poorly written applications will be vulnerable to the attack. The core components in Android use the proper method of querying and are therefore safe from the attack. However, any additional application installed on the device regardless of its source must be inspected to determine its vulnerability.

## 5.6 Connectivity and Communication

Multiple communication transports (Bluetooth, Wi-Fi, GPRS, UMTS, Cable) provide many options for malware to infiltrate a device. Some malware can propagate through more than one transport. For example, Lasco is a malware which spreads via the Bluetooth on Symbian devices [Emm, 2005]. In addition, it also infects all Symbian Installation Source (SIS)-files using social engineering.

Bluetooth on the G1 device supports pairing (i.e. two-way authentication between Bluetooth devices). In order to pair, one device out of the two must be set at "discoverable", which means that other devices may find out about its existence. Android allows itself to become discoverable, but only for a short duration of two minutes required for the pairing process. This significantly decreases the likelihood of being detected by attackers.

The USB protocol is entirely managed by software, and is therefore similar across Android devices. There are two sub-protocols which are supported by Android: mass-storage class (e.g. flash USB), and the Android Debugger Bridge (adb). The USB connection does not support the network, audio or other classes. By default, adb is disabled, and the computer regards the device as a mass-storage device, with no additional functions. Only the device's SD card through USB, not its system or data partitions are exposed. When "USB debugging" (adb) is enabled, the device can be managed with the same "adb" tool which is provided in the Android SDK. This tool makes it possible to push and pull files to and from the device, install apk files, TCP and UDP redirections, etc.

One of the Android-specific Linux kernel changes is the "Paranoid-Network". Usually, on Linux systems, a user-space process can open network connections at will. On Android, a user-space application must receive the "INTERNET" permission in order to make any kind of network connection. Because the change is at the kernel level, the application framework does not participate in enforcing these permissions. Indeed, even native applications are subject to this setting. The "Paranoid Network" setting works by hard-coding several POSIX group IDs in the kernel. An application must be a member of the relevant groups before it is allowed to create sockets.

### 5.7 Hardware

Hardware can be attacked as well. In the case of a cellular phone, several components are particularly vulnerable. Such cases are mainly relevant due to Android's tendency to allow applications to access as much of the platform as possible. Flash storage (SD card, internal and, SIM card) can be worn out since they have finite number of erase cycles and Android does not rate-limit I/O or allow any total I/O quotas. Draining the battery is trivial (e.g., by keeping the CPU running or by holding a wake-lock) and it is not easy to defend against such cases.

The T-Mobile G1 device does not provide an externally accessible JTAG connection. JTAG is a physical connection used for low-level debugging of digital systems and also for replacing operating systems. It can therefore be used to sidestep any software-only security mechanism. The term "jail-breaking" refers to the ability to install unofficial updates on a locked device. Assuming that the software forbids re-flashing and that no known vulnerability exists, "jail-breaking" will require physically opening the case and basic soldering expertise, which is more difficult than software-only jail-breaking.

### 5.8 Software Updates

Software update is a commonly used security mechanism which provides the ability to update the system with fixes to cope with lately discovered vulnerabilities. When connected to the Internet, the Android system can be pushed with software updates to fix known problems in vulnerable devices. Updates can be obtained over-the-air (OTA) after confirmation from the user or placed manually on the SD-card. The T-Mobile G1 device learns about OTA updates through periodic queries to a server over plain HTTP. During these requests for updates, private information (e.g., IMEI, Android ID, version number) is being exposed over unreliable links, and the user is left without the ability to disable it. The update file, signed by the provider, is verified in two stages. Firstly upon download (relevant only to updates obtained OTA) using public keys that come with the device and secondly by the recovery utility using additional public keys which are hardcoded into the executable. The system will not install an update which is not signed by an approved key. Therefore we conclude that the design of the software updating mechanism is sound, and assuming the implementation has no bugs, it is not possible to install an update which was not created by the owner of the update keys.

So far, the Android has had more than four updates (the latest of which is the CupCake version) which included fixes for coping with bugs, reducing vulnerability and for adding additional features. All versions up to and including RC29 had code enabled from a debugging phase that caused every key press on the local keyboard to be sent to a background root shell. This root shell vulnerability was used as the first "jail-breaking" method of Android devices and was subsequently fixed in RC30. When gaining root access on the device, a jail-break is capable of preventing any future OTA updates. Later versions closed this bug, therefore "jail-breaking" had to be done either by using known local privilege escalation attacks, or by firmware downgrades.

Firmware downgrade is a method that is used in order to gain root privileges on the device. If an earlier firmware version is known to be vulnerable and provides a method for gaining root privileges, a user/attacker will try to downgrade the firmware version to the vulnerable version. Android upgrade scripts employ version checks in order to prevent downgrades. For example, T-Mobile G1 does not allow RC29 (a vulnerable version) to be installed on a RC30 (or later) device. A way to sidestep this mechanism for the G1 was leaked in the form of a dreaimg.nbh file. Such a file represents a low level flashing method that only depends on the bootloader (and not on the code that does the normal .zip-based updates).

Another option for gaining root access is by using fastboot, a feature that allows the physical user to specify an alternative boot image. Such an image could for example an unofficial update that enables root access. Once it is gained, it is possible to permanently disable future updates. The only hindrance is that fastboot is usually disabled, as for example on the T-Mobile G1 and HTC Dream; however, HTC Magic has the fastboot "boot" option enabled.

### 5.9 Relevance of Existing Linux Malware

An attacker may attempt to target services, listening in on local host ports which would be inaccessible externally. These might include Android services such as the dbus (Bluetooth related Inter-Process Communications) or mountd (handling mounting file systems). Since some of these services run as the root user, any exploitable bug found related to them could allow an attacker to run his/her code along with root privileges.

The underlying kernel code is also a source of exploitable bugs. Any kernel vulnerability patched in kernel versions beyond 2.6.25 (Android's current kernel version) may still be applicable on Android. Kernel code security issues are likely to be more difficult to iron out due to the lag-time involved with issuing a large kernel patch to a wide user-base. This time delay may leave devices vulnerable for long periods.

Remote exploit attacks require communication with some piece of software running on the device. By default, none of the native services are listening for incoming connections. However, the "USB debugging" service is also listening for incoming TCP connections which are available over the network.

Root-kits, key-loggers and other types of malware belong to the post- exploitation phase and assume that there exists a connection enabling their transmission to the target system. They usually require root level access to the system by other means in order to install them. The rathole and Linux rootkit V (lrk5) root-kits, and vlogger key-logger were successfully compiled and activated on Android. The running root-kits on Android made it possible to remotely explore running processes, hide or kill specific processes, prevent hidden processes from being stopped, packet sniffing and provide several methods to communicate with the root-kit such as encrypted SSH backdoor and remote shell. Using the key-logger we managed to log keystrokes on the devices keyboard.

### 5.10 Relevance of Existing Java Malware

The Android platform does not support Java applets (due to licensing issues), or Adobe Flash and is therefore immune to most web-based Java exploits. However, such support may be added in the future and Google has already announced forthcoming support for Flash, potentially exposing the system to Dalvik-specific attacks over the web.

Pure "desktop" Java malware spreads by injecting code into other class files without harming the valid structure of the victim class file and its verifiability. The injected code can cause unexpected behavior. Java malware is not very widespread and only a few viruses been found since the distribution of the first two: StrangeBrew and BeanHive in the late Nineties [Plantey, 2005]. These viruses search the user's working directory for writable class files and then modify them to start execution from the viral segment.

Java viruses are not applicable to the Android framework for two reasons. First, they infect class file formats and must be adjusted to support injecting malicious code into Android binaries .dex files. Second, in Android, applications do not have write privileges to any Android package (.apk) files, even to their own. Moreover, since it is not possible to list a folder of another application for its files, any effective search for files to infect is not feasible. We modified and ran the StrangeBrew virus on a G1 phone and as we expected, it did not manage to access any .dex file or class file installed on the device.

Forcing Windows OSs (XP and Vista) to automatically run a malicious Windows executable from a G1 device (located on the SD-card) by using an Autorun.inf file was tested and found not feasible. Only the icon that represents the device was changed according to our Autorun.inf settings. It may be possible to exploit the fact that the icon is being loaded by the system to trigger a buffer overflow using a malformed icon file.

The autorun capabilities are restricted to CD-ROM drives and fixed disk drives. USB storage device can perform autorun if it contains an autorun.inf file and a startup application, and most importantly, not marked as removable media device. The SD-card in Android is marked as removable disk. When connecting the Android device to a PC, it will appear as an additional drive, but the drive will not be available until after mounting. After which the Windows system will not be instructed to run the autorun.

## 6. SECURITY ASSESSMENT OF THE ANDROID FRAMEWORK

Based on our security analysis, described in previous section, we derived the following conclusions [Shabtai, 2009d]. First, an Android device in its normal state is well-guarded since neither the core components nor the kernel can be replaced by an attacker (or by the legitimate owner), unless the hardware has been manipulated, which is difficult to perform. The only way to alter operating system components is to identify a vulnerability in one of the kernel modules or core libraries that will enable the virus to acquire root access.

Whenever publishing a bug or vulnerability in one of the core components (such as a native library or kernel component), an attacker might be able to run malicious code in a highly privileged mode and even gain full control over the device. This threat is amplified due to the fact that Android's source-code is publicly available; some system processes run with root privileges; and no fine-grained access control mechanism exists for system processes. We believe that this scenario is likely to be realized. Making the source code publicly available provides certain benefits, among them, many individuals will be able to check and verify the code. Even if the platform is never fully secure, the open source approach promotes steady, ongoing security improvements. As time passes, we expect the number of bugs to diminish and the system to become less vulnerable.

In order to remotely attack an Android device, it will have to expose a vulnerable service to the Internet. This requirement reduces the likelihood of remote attack scenarios since by default none of the services are listening for incoming connections. The amount of exploitable code running on a device through local services, device drivers and kernel code makes host-based exploitation attempts a higher risk. Thus, we regard the device more vulnerable to local host-based exploitation attempts.

The permission mechanism is not sufficiently protected and installation of an application that maliciously uses permissions granted by the unaware user is a scenario which is likely to occur. The framework also provides the adb install feature that makes it possible to install applications and to grant permission to an application without any user interaction. In addition, a user cannot approve a sub-set of requested permissions (it is "all-or-none") and cannot verify that an application uses its granted permissions only for benign purposes. Moreover, the shared user ID mechanism allows sharing permissions between applications without a user's awareness or the need for explicit approval.

Malicious code injection via a Web browser to be possible. WebKit, Android's open source web engine, has a history of vulnerabilities. Some recent attacks on the Web browser include a buffer overflow in an outdated native library, and an explicit XSS vulnerability. Both attacks resulted in the attacker being able to run any malicious code on the device with the abilities and privileges assigned to the web browser application.

Injecting malicious applications via Bluetooth is, however, not likely to occur as there are several protection mechanisms: (1) a device can set the Bluetooth connection as not discoverable; (2) if the Bluetooth connection is set as discoverable, it is only for a short period of two minutes; (3) the owner needs to accept the connection; and (4) the owner needs to manually install the file.

Our security inspection indicated that the system is well protected against SQL injection attacks. However, some information is fully exposed to attackers (for example, all the content on the SD-card). Separation of user IDs between applications protects the device and other services and applications from tampering.

Figure 2 presents the results of a qualitative risk analysis that we conducted in order to identify and prioritize the threats that an Android-device may be exposed to. The evaluation is based on assessing the impact and likelihood of various threats (listed in Appendix A[4]) exploiting vulnerabilities (listed in Appendix B) in Android in order to harm, disable or abuse the confidentiality and/or availability and/or integrity of the following assets of the Android framework [Dagon, 2004]: (1) *Private/confidential content* that is stored on the device (pictures, contacts, emails, documents etc.); (2) *Applications and services* (phone, messaging, emailing, Internet); (3) *Resources* such as the battery power, communication, memory and processing power (CPU); and (4) *Hardware* includes the device itself, external memory card, battery and camera. Our mapping of the most important threats was done into three levels of likelihood: Unlikely, Possible, and Likely; as well as three optional Impact values: Minor, Moderate and Severe.

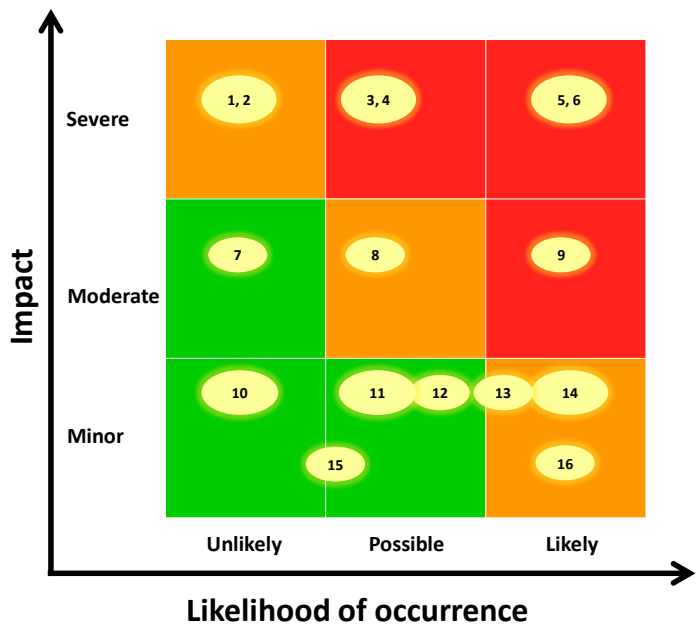

Fig. 2. Qualitative risk analysis results (listed in Appendix A)

---

[4] Threats 17 and 18 were omitted since they are beyond the scope of this paper

Following the above conclusions and the risk assessment, we identified the five most important threat clusters which should be countered by employing proper security solutions/capabilities. These threat clusters were obtained by grouping similar threats assigned with the highest risk:

**Threat cluster-1**:
> **Compromising availability, confidentiality and/or integrity by maliciously using the permissions granted to an installed application**. This attack scenario is likely to happen and potentially has a high impact on the device.

**Threat cluster-2**:
> **Compromising availability, confidentiality and/or integrity by an application exploiting a vulnerability in the Linux kernel or system libraries**. This scenario was proven possible and our security analysis shows that additional vulnerabilities are likely to be found. Although, it has a low probability of occurring, it carries a potential to inflict severe damage.

**Threat cluster-3**:
> **Compromising availability, confidentiality and/or integrity of private/confidential content**. Contents on the SD card are not protected by any access control mechanism. Additionally, wireless communication can be eavesdropped remotely.

**Threat cluster-4**:
> **Draining resources**. There is neither disk storage nor memory (RAM) quota per application. Hogging the CPU is also possible.

**Threat cluster-5**:
> **Compromising of an internal/protected network**. Android devices can be used to attack other devices, computers or networks by running network or port scanners, SMS/MMS/email worms and various other methods of attack.

## 7. APPLICABLE SECURITY MECHANISMS FOR ANDROID

Several security companies have already announced their intention of modifying their security solutions for Android. These solutions include various antivirus software and intrusion detection systems that run on the devices. As a case in point *SMobile* announced on November 2007 of their security solution for handsets based on the Android SDK[5]. Their solution, "Security Shield" includes antivirus, anti-spam and firewall applications. *Savant Protection*, which specializes in intrusion prevention, announced on March 2008 that its security solution "Savant Technology" was ported to Android[6]. *Mocana* is another security company that ported its solution to Android, and claims to provide the following capabilities[7]: a secure browser, virtual private network (VPN) clients to secure data communications between the device and corporate network; voice, video and data encryption; malware and virus protection; scalable and secure firmware updating and secure boot capabilities; and robust certificate handling features to authenticate devices, network services, and individuals to each other. DroidHunter is another security application available for Android. We expect that other security companies, which already have security solutions for smart phones (such as Symantec, F-Secure, McAfee and CheckPoint) will also port their solutions to Android-based devices.

In order to further harden an Android device and mitigate the identified high-risk threats several additional safeguards may be employed. Some of these mechanisms were tested and evaluated in our mobile security laboratory. In the following paragraphs we provide a description of several security mechanisms which can be adapted to harden the Android. These mechanisms include tools that are already ported to Android, Linux portable security solutions and feasible, developable solutions. The arsenal of security mechanisms is summarized in Table III. Each of the potential countermeasures is described with its strengths and weaknesses, realization approach and estimated level of effort required for implementation.

---

[5] http://secure.smobilesystems.com/main/home/index.php
[6] http://findarticles.com/p/articles/mi_m0EIN/is_2008_March_11/ai_n24384617
[7] http://mocana.com/NanoPhone-Android.html

When considering the applicability of a security measure, we tried to determine who would be implementing it and what should be the realization approach. We cluster Android security measures into three types in terms of their realization approach:

- **System modifications** – Requires altering Android's core source-code including the framework, Dalvik VM, Linux Kernel, daemons, native libraries, etc. Also, we have chosen to put in this category any course of action that requires modification of Linux kernel configuration, because of the close interaction with the rest of the system. The prime advantage of system modifications is the ability to add new functionalities to the system and change its existing behavior. The prime disadvantage of system modifications is that it is relatively expensive in terms of man power and time, requires rigorous testing and validation.
- **System add-on** – Requires modifications of Android's core configuration files (e.g., init.rc script); addition of applications that require Signature- or SignatureOrSystem-level permissions; or replacement of applications in the system image. The System add-ons approach, provides higher privileges either through additional Android permissions (Signature, System) or by running as root (e.g., full file system access, applying firewall policy, managing memory quota, listening to keystrokes/touch screen). In addition system image applications cannot be uninstall; such a safeguard may prove crucial for some security applications. The main disadvantage of System add-on applications is that they consume system resources which are quite limited, and that they make updates harder as the application needs to be re-installed.
- **Add-on application** – Can be applied by any user by simply installing an Android application. The add-on application is in the same status level as other applications, and consequently, may request the exact same permissions as any other application. Add-on security applications enable simple portability to any Android device and facilitate simplicity of maintenance. Although, most of the functionalities required for the security application are not available, and the limited capabilities of policy-enforcement (e.g., quota management), as the user can remove the application.

Table III. Applicable security mechanisms.

| Mechanism | Description | Security issue | Existing tools |
|---|---|---|---|
| **Anti-malware** | Scans files, memory, SMS, MMS, emails, URLs, Java scripts | Viruses, Trojans, worms, root-kits and other malware | SMobile, Mocana, DroidHunter, ClamAV [Shmidt, 2008] |
| | **Realization**: System add-on | **Effort**: Low | |
| | **Pros**: Well-known; extensively used in other platforms; low false positives. **Cons**: Detects only known malware; not effective at this time on mobile devices; requires frequent updates. | | |

| Mechanism | Description | Security issue | Existing tools |
|---|---|---|---|
| **Firewall** | Can block and/or audit un-allowed connections from/to the device | Services which are exposed to an un-trusted network; prevent network attacks | SMobile, Netfilter/iptables |
| | **Realization**: System modification | **Effort**: Low | |
| | **Pros**: Well-known; extensively used in other platforms; highly effective. **Cons**: Will not protect against attacks on browser, email, Bluetooth, SMS/MMS. | | |
| **IDS/IPS** | Detects abnormal or known malicious behavior of the system, process, network traffic or user | Frauds (e.g., expensive calls), unusual telephone activity, theft, malicious attacks | Andromaly [Shabtai, 2009a; Shabtai 2009e], DroidHunter |
| | **Realization**: System modification, System add-on | **Effort**: Medium | |
| | **Pros**: May detect new and isolated attacks; may be adapted easily for any task. **Cons**: May consume high resources; high miss rate and false positive. | | |
| **Access Control** | Limiting access of processes and user to resources and/or services. | Limitation of damage from malicious/exploited applications | SELinux [Shabtai, 2009c]; DiffUser [Ni, 2009] |
| | **Realization**: System modification | **Effort**: Low | |
| | **Pros**: Low resource consumption; provides effective protection for high privileged processes. **Cons**: Hard to configure; deployment of modified kernel. | | |
| **Login** | User should provide a secret to use the device | Unauthorized use of the device | Android screen lock pattern |
| | **Realization**: System modification | **Effort**: Low | |
| | **Pros**: Widely used on other platforms; known to users; effective when the device is stolen. **Cons**: Might be annoying to the user; does not protect all information; passwords/secrets may be broken eventually. | | |
| **Selective Android permissions** | Allowing a user to grant only a subset of permissions to; limiting permissions using predefined policy; relevant mainly to corporate users | Protecting from granting un-needed permissions that can be maliciously exploited | SAINT [Ongtang, 2009], Kirin [Enck, 2009a] |
| | **Realization**: System modification | **Effort**: Medium | |
| | **Pros**: Effectively protecting from granting sensitive permissions to malware. **Cons**: Requires changes in design; applications may crash; only for advanced users, policy maintenance is costly. | | |
| **Permissions management application** | Scans Android's applications' permissions, presenting the user a concise summary | Installed unwanted applications, Trojans | |
| | **Realization**: Add-on application | **Effort**: Low | |
| | **Pros**: Can detect installations of unknown applications; partially overcomes the SharedUID issue; can be implemented by the HIDS developed in this project. **Cons**: The user is still left with the decision of what to do with the application: use it or uninstall. | | |

| Mechanism | Description | Security issue | Existing tools |
|---|---|---|---|
| **Data encryption** | Encrypts private content on the device | Protects sensitive information in case of theft or against malicious application | Enabled on Android release 1.5 (i.e., Cupcake) |
| | **Realization**: System add-on | **Effort**: Low | |
| | **Pros**: Known effective algorithms; the only applicable solution for protecting sensitive data on the SD-card.<br>**Cons**: Might be annoying to the user; to support encryption of SMS, emails and contacts requires system modification. | | |
| **Phone call encryption** | Providing secured connection (authenticated, encrypted) | Eavesdropping, identity verification | |
| | **Realization**: Add-on application | **Effort**: Medium | |
| | **Pros**: Effective against eavesdropping.<br>**Cons**: Encrypting phone calls requires both parties to share the same application; reduce the quality of calls; not commonly used. | | |
| **VPN** | Connects to a remote network over the Internet; relevant mainly to corporate users | Provides secured connection to networks | PPTP, L2TP and IPSec-based VPN connections are enabled on Android release 1.6 (i,e., Donut) |
| | **Realization**: System add-on | **Effort**: Low | |
| | **Pros**: Effective and widely used mechanism; adopted by many companies.<br>**Cons**: Increases network latency. | | |
| **Spam-Filter** | Blocking MMS, SMS, emails, and calls from unwanted origin | Spam | |
| | **Realization**: System add-on, Add-on application | **Effort**: Low | |
| | **Pros**: Known and effective technology; decreases negative user experience.<br>**Cons**: Might consume high resources; high miss rate and false positive; requires constant updating of filters. | | |
| **Application certification** | Each application should be signed by a Certificate Authority (CA) | Limiting potential damage caused by un-trusted applications | |
| | **Realization**: System modification | **Effort**: High | |
| | **Pros**: Highly effective in providing a mechanism that evaluate the nature of an application.<br>**Cons**: Expensive and hard to implement. | | |
| **Resource management** | Fairness in resource allocation (CPU for phone application, disk quotas, I/O rate limiting and quotas, network quotas, and traffic shaping) | Denial of Service (DoS) | |
| | **Realization**: System modification | **Effort**: High | |
| | **Pros**: Can effectively protect important applications and services.<br>**Cons**: Hard to implement and requires modification in the framework. | | |

| Mechanism | Description | Security issue | Existing tools |
|---|---|---|---|
| **Remote management** | Remotely configure and manage the device (settings, firewall policy, remote "bricking", apps tracking); relevant mainly to corporate users | Provide updated security, remote disabling (anti-theft protection) | |
| | **Realization**: System modification | **Effort**: Medium | |
| | **Pros**: Can provide a centralized solution for protecting sensitive data; detecting intrusions; keeping up-to-date security mechanisms. | | |
| | **Cons**: Requires human support and high maintenance, costly. | | |
| **Context aware access control** | Dynamically allow and restrict access to resources and services based on a pre-defined model | Protect confidential content and the integrity of services | Locale app in the Market |
| | **Realization**: System modification | **Effort**: Medium | |
| | **Pros**: Can protect the user without his/her intervention. Cons: Hard to define/learn policies. | | |
| **Integrity Checking** | Verification of system and application state | Offline tampering | |
| | **Realization**: System modification | **Effort**: Medium | |
| | **Pros**: Can protect the user without direct intervention. **Cons**: Hard to define/learn policies. | | |
| **Automated Static Analysis and Code Verification** | Analysis and verification of applications and data flow | Limiting potential damage caused by un-safe applications | Static analysis [Shabtai, 2009b; Shmidt, 2008]; ScanDroid [Adam, 2009] |
| | **Realization**: System modification | **Effort**: Medium | |
| | **Pros**: Can be highly effective in providing a mechanism that evaluate the nature of applications; pinpointing suspicious applications; and assisting in rapid certificaition of applications. | | |
| | **Cons**: Hard to define the indicators of malicious applications. | | |

## 7.1 Anti-Malware

To identify and remove malware, anti-malware software for mobile devices examines all files in specified locations, email attachments, the memory, system configuration, MMS, Bluetooth objects and other relevant areas. It usually identifies and exterminates known malware based on a signature repository.

As described earlier, several commercial solutions are available for Android which also provides an anti-malware component. There are also open-source antiviruses and rootkit detectors that can be ported to Android such as the ClamAV[8]. ClamAV is available for Linux-like operating systems under the terms of the GNU GPL, and the signatures are available free of charge. Usually running as a daemon and serving inspection requests by other processes, it is mostly useful as an email attachment scanner and network file server/proxy scanner. Porting ClamAV into Android requires the following general steps: compiling ClamAV to ARM with the bionic C library; writing GUI for ClamAV specifically to Android; integrating on demand scanning with Android's browser and mail application; and finally preparing relevant signature repository. An attempt to port ClamAV to Android was presented by Schmidt et al. (2008).

Anti-malware is a well-known solution and is extensively used in other platforms. Signature-based solutions provide low false-positives, but will only detect known malware and require continuous updating of the signature repository. At this time, the anti-malware solution does not seem to be effective for mobile devices.

---

[8] http://www.clamav.net/

## 7.2 Firewalls

A firewall running on Android-powered devices can limit the vulnerability to remote network attacks by preventing Internet-based scanning and access to internal services.
The SMobile solution that was ported to Android also claims to include a firewall tool. Linux 2.6 includes the iptables integral firewall which uses the NetFilter framework. NetFilter is a Linux kernel subsystem that provides packet filtering, rewriting and connection tracking capabilities that are used to implement firewalls. Capabilities of iptables/NetFilter and associated modules include: filtering incoming, outgoing or forwarded packet; matching packets by rule-matching on protocol fields (e.g., IP protocol, src/dest address); stateful inspection; and dropping, rejecting or accepting a packet with an ICMP notification.
NetFilter is enabled on the T-Mobile G1, thus only a control application is needed. Due to the requirement of permissions in order to update firewalling policy, the control application could be implemented as a system add-on. We have developed a preliminary control GUI. It depends on the "su" application from common unofficial firmware. "su" is a program which allows running commands as "root", and in our case is used to run the iptables command with the required super-user privileges. From our experiment we can classify the effort needed to develop such an application as low.
While a firewall is a well known and highly effective solution that may be extensively used in other platforms, it will not protect against attacks via browser, SMS/MMS, email or Bluetooth and will not provide phone call filtering.
In the basic firewall that we activate on Android rules are very simple and provide the ability to block/deny communication to specific IP address/DNS/ports. The more suitable firewall policy for Android would be similar to the MS Windows firewall which allows defining rules at the application level. In such a way we can define for each application who can access it and which application can send information and to where. We can make sure that port scanning is not preformed from the device by a malicious application, block access to specific websites and more.
Firewalling at the application level on Android is not a trivial task since any application that is granted with INTERNET permission can open a socket at its will. Thus there is a need to catch an application that tries to send/receive data and apply the policy. The application permission mechanism is not fine-tuned enough and thus we see a firewall as a solution that can increase security and overcome the problems in the application permission mechanism.

## 7.3 Intrusion Detection/Prevention System (IDS/IPS)

IDS/IPS capabilities tailored for smartphones have been offered in recent years by leading security vendors. As a case in point, Trend Micro Mobile Security[9] claims its products defend against malware and help prevent unwanted intrusions and data leakage through new firewall and intrusion detection technology. Norton Smartphone Security by Symantec provides IDS capabilities for Symbian and Windows-Mobile phone[10].
The Linux Intrusion Detection System (LIDS)[11] is an enhancement for the Linux kernel (implemented as a kernel patch). It implements several security features such as

---

[9] http://us.trendmicro.com/us/products/mobile-security/index.html
[10] http://www.symantec.com/norton/smartphone-security
[11] www.lids.org

Mandatory Access Control (MAC), a port scan detector, file protection and process protection (restricting even root). OSSEC HIDS is an open source intrusion detection system for Linux that performs log analysis, file integrity checking, root-kit detection, time-based alerting and active response.

As stated earlier, most academic initiatives to enhance protection of mobile devices have employed host-based intrusion detection systems comprising an agent collecting various features from the device and then applying various machine learning algorithms to classify the behavior of the system as benign or malicious or to detect anomalies.

Host-based intrusion detection/prevention systems (HIDS/HIPS) monitor devices, applications or the user to detect abnormal or known malicious behavior. When an unapproved action or abnormal behavior is performed, an action can be taken to notify or correct the situation. It can be used for user verification by monitoring the activity of the user on the device (which application is used and in what sequence, keyboard usage - typing rate, touch screen etc.) and by verifying that the logged-on user is the same as the authorized user. Anomaly-based IDS can detect unusual telephone activity, malicious attacks such as denial of service, and protect the information on the device in case of theft or loss. While it may detect new and isolated attacks, it will probably suffer from high rate of false positives and a certain miss ratio.

In our Android security research we developed and evaluated the "Andromaly", which is an experimental anomaly-based IDS for Android [Shabtai, 2009a; Shabtai, 2009e]. Andromaly employs various methods, such as anomaly detection and temporal reasoning, to facilitate detection of maliciously behaving applications. Based on our experimental implementation, we conclude that the development effort is medium and that our IDS requires system add-ons for basic capabilities and system modification for more advanced capabilities.

As stated by Rich Cannings[12], Google's Android Security Leader, the Android Market was chosen to be the place for reporting security issues by users, and users can mark applications as harmful, which trigger the security team to launch an investigation as in the case of the MemoryUp application[13]. An Intrusion Detection System such as the Andromaly can be used for reporting suspicious behavior of applications to the Android Market.

## 7.4 Access Control

Android incorporates several access control mechanisms. While these mechanisms are enforced on the application level or only on files, Linux can provide other tools that are directly enforced by the kernel. One framework, the Linux Security Modules (LSM) that allows the Linux kernel to support a variety of access control models. Zhang et al. (2007) introduced the trusted computing and SELinux into mobile phones.

We tested on an Android G1 device the Security-Enhanced Linux (SELinux), which is the best-known implementation of a Linux Security Module (LSM) [Shabtai, 2009c]. SELinux allows restricting of any process in the system, including root-owned, and by that limiting access of processes and user to resources and/or services, thus limiting the potential damage from malicious or exploited applications. Its decisions are based on an access control policy, which should be deployed together with the base system.

The main issue that SELinux seeks to solve is confinement of system processes, particularly those with high-privileges. Since all system processes are known a priori, the policy would specify exactly what legal actions each service can perform. Should a vulnerability be found in one of these services the SELinux confines the process and thus limits the attacker's maneuverability. SELinux is regarded as a preventive measure.

---

[12] http://www.usenix.org/events/sec09/tech/tech.html#cannings
[13] http://www.geek.com/articles/mobile/google-says-that-memoryup-has-no-malware-20090128/

In general, all normal activities on the device should be approved by the SELinux policy, and any denial during normal operation should be considered as a bug. Moreover, there is no user interaction, and therefore no usability impact. The SELinux policy is hard to derive and configure while its integration of SELinux requires deployment of modified kernel (i.e., system modification). Our experimentation with SELinux on Android has shown that it consumes very few resources and incurs a very low overhead [Shabtai, 2009c].

Ni et al. (2009) describe DiffUser, a framework for providing access control mechanism on smartphone for different users; more specifically, administrator, normal user and a guest user. DiffUser was implemented and evaluated on Android. Each user can be assigned with different rights e.g., only administrator can install/uninstall applications; the guest user can only use the phone application.

### 7.5 Owner Login

In private and public computer networks (including the Internet), authentication is commonly done through the use of log-on passwords. A similar mechanism can be developed for Android. When powering on a device, the user would be required to enter a secret (e.g., password, or a shape using the touchscreen) as specified during the device setup and known only to the user. The device can be re-locked after several wrong attempts or, automatically after an inactivity period. In order to use it again the owner will be required to re-enter the secret. An additional feature to protect sensitive data would be locking-up the device by means of a special SMS that the user can send in case the device is stolen. Biometric means for identification (i.e., fingerprint scanning or using a smartphone's camera) are also beginning to emerge for mobile devices.

Although, these secret-based solutions might be annoying to the user (the need to provide the secret for every use), it may not protect all information (for example, the data on the external SD-card will be still exposed) and passwords/secrets can be broken. At present, Android is provided with a simple screen lock pattern mechanism.

### 7.6 Protecting Android Permissions

During the installation of application on Android, the user may view a list of required permissions, and may decline installation based on this list. There is no way for the user to allow only a subset of the required permissions. In practice, the user is unlikely to deny installation of an application he or she wants based on such a list. We describe two optional mechanisms aimed at tightening access to Android Permissions.

*7.6.1 Selective Android Permissions*

Sometimes permissions are requested only for esoteric features by the application (e.g., a game may request "Internet" access for uploading high scores). Sometimes, the permission is requested for a valid use case, but the user does not plan to use this feature. For example, the ChompSMS allows sending SMS using both the mobile carrier, and via the Internet, but a user may wish to use only one method.

We propose adding an advanced feature to the Package Installer enabling the user to decline certain requested permissions but still permit installing of the application. Such a change would be highly beneficial to security aware users, but would not impede usability for unaware users. This solution would protect from granting unneeded permissions that could be maliciously used. Overall, the required effort is low, but requires a system modification and possible changes in the design. In addition, applications granted with a partial set of permissions may crash if the developer did not anticipate and provide a solution for such a situation (i.e., handle cases in which partial permissions were given).

This solution can be enhanced to provide the option for hardening Android devices by limiting granted permissions using a predefined policy. The goal is to protect users from granting unneeded permissions that can be maliciously used. It is mainly relevant to corporate users. Enterprises can harden the Android devices of their employees by restricting certain permissions from being granted on such devices. When an application is about to be installed, the Package Installer would compare the requested permissions with the policy and would not grant permissions that are being blacklisted. The policy would be defined by the enterprise's IT personnel who grant the device to the employee.
A sophisticated policy could blacklist a combination of features, allow certain permissions only to selected application signers, limit resources, etc. A policy could be deployed as a file or be available online for reference when necessary.

Efforts for enhancing Android security at the application level permissions are presented by the Kirin system [Enck, 2009a] and Secure Application INTeraction (SAINT) [Ongtang, 2009]. These two systems presented an installer and security framework that realize an overlay layer on top of Android's standard application permission mechanism. This layer allows applications to exert fine-grained control over the assignment of permissions through explicit policies. Since implementation of such a mechanism requires changing the Package Manager, it is regarded as a system modification.

*7.6.2 Permission Management Application*

This solution periodically scans Android's application permissions and files world_read/write_access in order to detect installations of unknown applications. This solution partially overcomes the shared user ID issue. The motivation for an Android permission scanner would be to assist the user in detecting applications with undesired permissions (which perhaps were accidentally granted at install time). Another aspect is the containment of a security hole caused by the existence of files which can be accessed and/or modified by all applications (global read/write permissions). In order to remedy such a situation, a permission scanner would present the user with permissions granted to each of the applications installed on the device. With the aid of such a display, odd permissions would pop right up. For example, a media player that has the permission to listen to keyboard events should catch the user's attention. If a permission anomaly is detected, several courses of action are possible: simply revoking the undesired permissions; uninstalling the application; blocking the application from running; or requiring user-approval when the permission is actually used by the application, thereby facilitating one-shot, or one-session approval. All possible actions except the first will require major changes in the framework. When using the first action, the user would still have to decide what to do with the application: either use it or uninstall it. Such an application could be simply implemented with low effort as an add-on application.

7.7 Data Encryption

A mobile phone that is lost or stolen might contain important personal contact data, corporate e-mails, or vital and confidential company data. In order to protect sensitive information, stored on the device, data can be encrypted. Removable media cards that are plugged into the devices can also be encrypted. The encryption is password-based and in order to access the device or data on the device, users authenticate with a password or PIN. Policy controls such as limiting the number of password retries can be implemented for greater security. Encryption capabilities were merged into the Cupcake update.

### 7.8 Phone Call Encryption

Phone call encryption could provide secured voice connection by means of authentication and encryption, and would effectively target eavesdropping phone conversations. Phone call encryption requires both parties to share the same application and it may reduce the quality of calls. It is mainly used for military purposes. Using the permissions provided by the Android application framework, this solution could be provided as an add-on application.

### 7.9 Virtual Private Network (VPN)

A VPN solution is relevant mainly to corporate users and effectively provides a secured connection to protected/private networks over the Internet. Some VPN tools for Linux exist, with OpenVPN being the leading, open-source contestant. Connecting to Microsoft servers over Point-to-Point Transfer Protocol (PPTP) is possible using PPTP Client. VPN can also be deployed using OpenSSH (version 4 or later) over the SSH protocol. Linux also supports IPSec, which is a standard-based method for establishing secured connection which is also used for VPNs. PPTP, L2TP and IPSec-based VPN connections are enabled on Android release 1.6 (i,e., Donut update). Enabling additional Linux-based VPN solutions on Android involves only a low effort. The required root privileges for creating a virtual network adapter makes it a system add-on.

### 7.10 Spam-Filter

A spam filter blocks unwanted MMS, SMS, emails, and calls from an unreliable origin. Previous industry experience from e-mail spam indicates that there are two prominent methods of dealing with spam. The white/black listing approach where sources are either known as good (white-list) and therefore delivered, or as bad (black-list) and therefore blocked. A second approach is having the system filter incoming items using machine learning (ML) classifier filters, based on features from the transport system and/or content of the message. The classifiers are trained and updated over time. The ML approach can also use the time of day and a user profile for a more fine-grained classification of incoming items. In the mobile phone arena, the white and black list approach is more common, with Caller ID being used as the source for allowing/blocking a call. SMobile's solution that was ported to Android, includes the PointGuard solution that can filter spam messages and calls based on updated blacklists. Kaspersky Mobile Anti-Virus 6.0 provides anti-spam capabilities that block telephone numbers of known spam sources, incorrect numbers or unwanted words or phrases that were added to a blacklist. It also supports a white list. Norton Smartphone Security and TrendMicro Mobile Security 3.0 include an SMS anti-spam protection feature that can block short messages (SMS and MMS) from unknown senders.
Since a spam-filter on mobile devices might consume high resources and may suffer from a high miss rate and false positives, it requires continuous updating of filters.
An anti-spam solution could be developed for the Android and employ both white/black-listing and the ML method in order to filter spam SMS, MMS, emails or phone calls.

When using ML classification methods, phone calls could be ranked with the belief that the call is in fact a spam call allowing the phone owner decide whether he or she wishes to take the call or not. Since G-Mail, on which Android heavily depends, already provides server-side spam filtering and connection to other mail providers, an anti-spam solution would provide significant advantage. Since the mail application does not require any special permission, and there are other free e-mail clients for Android, realization could be a system modification.

### 7.11 Application Certification

An implemented solution by other mobile operating systems (i.e., Symbian), and advocated in the OMTP Application Security Framework [OMTP, 2008] has different trust levels for installed applications. An application that is installed on the mobile device is allowed to request different permissions (e.g., initiate an outgoing call, create a network data connection using HTTP/HTTPS, send SMS or MMS, determine the current location of the device using GPS) depending on its trust level. The trust level is assigned to the application according to its origin using a signature and third-party certification mechanisms. Whenever there is an attempt to install an application, the first step is to validate its certificate. If there is no certificate or the validation of the certificate has failed, the application is assigned the lowest trust level and can only be installed while requesting the basic harmless permissions. Alternatively, the installation can be always aborted. If the certificate is valid, the permissions requested by the installed application will be granted or denied based on its associated trust level. If for instance the application requests permission to access the Internet but its trust level does not permit it, said permission will not be granted and will cause a run time error if an API requiring it is used.

Android uses certificates in a limited way in order to ensure package integrity and that two or more packages are from the same origin. Applications that define their own permissions may choose to grant such permissions only to packages sign by the same author. There is no support for root Certificate Authorities (CAs) or for certificate chains in Android. In order to employ the application trust mechanism, Android should be modified to support trust levels of applications, associating CA certificates to the trust level, as well as verifying certificate chains. This mechanism is highly effective in evaluating the nature of an application, detecting malicious applications before they are installed on a device and limiting any potential damage caused by untrusted applications. However, this solution is highly expensive in terms of implementation and maintenance.

Although certification has been proven very effective, it is not error prone and malicious applications can still unintentionally be approved and signed. In addition we can assume that users will continue to download and install "unapproved" applications that are available from free Websites and prefer them over trusted applications that need to be paid for. Furthermore, Android is grounded in an open source approach, while the certification framework contradicts this approach; thus researchers should look for alternatives to capture application semantics without relying on manual code inspection. One approach which is closely related is static analysis and verification of code (see section 7.16).

### 7.12 Resources Management

There are four main resources in a computing environment such as Android: CPU, storage, RAM memory and I/O. Each application can request as much of each resource as it wants. Not all of these resources have safeguards against unfair usage.

**CPU**: Android employs Linux's Completely Fair Scheduler (CFS) that ensures that an equal share of CPU is distributed among all processes. In addition, specific processes can be granted a larger share, but are still prevented from monopolizing the CPU. To test this fairness mechanism we have created a simple application that starts 100 threads that loop doing nothing in particular. Running this application on a G1 device resulted in the entire device being frozen.

**Storage**: Linux supports storage quota, but Android does not enable it. This means that currently each application can create as large files as it wants, both on the internal flash, and on the SD card. The application list shows the sizes of each application, so that one is able to uninstall large applications, but that is not enough; files can be created outside of the private application folder and therefore do not counted in the application list figures. We suggest storage quotas in the default configuration.

**RAM**: When there is not enough free memory in the system, Linux kills some process to free RAM. Android has added an even more aggressive lowmemorykiller for the same task. In practice, it has been demonstrated that this killing process can be invoked by visiting a malicious website, such that all applications on the system will be killed, including core services such as the system server. When the system server is killed, the device restarts. That aggressive behavior may be modified.

**I/O**: Both network, and disk I/O bandwidth are limited. There are several I/O bandwidth controllers for Linux, but none are activated on Android. Another network example in this category is called "traffic shaping", but it is also disabled in Android.

In regard to some applications which are considered critical, such as the "phone" application, we suggest tuning safeguards to grant larger slices to such applications. All of the resources mentioned support defining larger slices for specific processes. All of the resource management configurations that were mentioned require system modifications. Due to the invasiveness of such a change, the effort involved in implementation scales from medium to high, depending on the exact implementation (enabling a configuration, or adjusting the framework to provide support for such implementation).

### 7.13 Remote Management

Remote management mechanisms consolidate several other security mechanisms while providing the ability to remotely control, configure and manage the device. This may include: remotely setting various parameters (e.g., Wi-Fi or Bluetooth network configurations); updating firewall policy; pushing security updates; updating anti-malware and anti-spam tools; tracking the device location; uninstalling/installing applications; remotely bricking the device; deleting or encrypting data; providing the means for remote assistance when a problem is encountered; and more. This solution is mainly relevant to corporate users.

Remote management of Android devices provides a centralized solution for protecting sensitive data in case of loss or theft, detecting intrusions and ensuring up-to-date security. It requires human support and high maintenance and thus is very costly. Due to high availability standards required by the enterprise market, the effort involved in implementation is medium. The required privileges for access to the relevant information, and effective control over the device would mandate a system modification.

### 7.14 Context-Aware Security

A context-aware solution can dynamically allow and restrict access to resources (documents, emails) and services (camera, Internet, phone, messaging) based on a predefined policy and on the instantaneous context of the device. It can provide better protection for confidential content and to ensure the integrity of various services. Context-aware applications, such as the Locale application in the Android Market, are already available, but they require high interaction with the user in defining the rules and are not security oriented. The challenge inherent in these solutions is to automatically learn and define policies, preferably without straining the user. Its implementation effort can be classified as medium and would require a system modification.

### 7.15 Integrity Checking

Integrity checking is used to validate that the system was not tampered with. It can prevent information exposure in scenarios such as replacing the phone application with a similar application that contains a Trojan that logs phone calls or eavesdrops on them.
The first approach for integrity checking involves monitoring changes in the file system in relation to a base-line state. Tripwire[14] is an open source integrity check utility that typifies this approach.
The second approach employs a hardware chip, and is geared towards ensuring system validity to a remote party, such as the telephony or content providers. As a case in point, Integrity Measurement Architecture (IMA), IBM's implementation of the Trusted Computing Framework for Linux, intercepts every file access/execution, and verifies its integrity (e.g. cryptographic hash) by employing a Trusted Platform Module (TPM) hardware chip (which is protected against the system software) of the measured system. This type of integrity measurement can be implemented in Android in several ways. Remote attestation allows enterprises to remotely check that their employee phones were not tampered with. It can make subverting the phone more difficult, if secure booting is implemented (although this is not in the openness spirit of Android). It can allow the user to encrypt files that will only be available on a specific phone.

### 7.16 Automated Static Analysis and Code Verification

An additional possible countermeasure can be provided by applying automated code analysis and verification. We took that avenue by exploring the use of machine learning classifiers on static features extracted from Android's application files [Shabtai, 2009b].

Android apk files encapsulate valuable information that can help in understanding an application's behavior. This information includes requested permissions, framework methods called by the application, framework classes used by the application, User Interface widgets and more. Using our apk feature extractor, we extracted features from the files including: (1) *apk features* such as apk size, number of zip entries and number of files for each file type; (2) *XML features* such as number of xml elements, attributes, namespaces, distinct strings and used permissions in the Android Manifest; and (3) *dex features* such as a Boolean for each method in the framework (used or not), a Boolean for every type in framework (type is used or not) for example MotionEvent, execSQL() etc.
Android uses a proprietary format for Java bytecode called dex. In order to extract meaningful features from dex files we reverse engineered the dex file parser, which is embedded in the Dalvik VM only, and developed a "dex" file parser. This parser can transform contents of the dex file into standard features (e.g., strings, types, classes, prototypes, methods, fields, annotations, static values, inheritance, modifiers, opcodes) and assisted in manually verifying the capabilities of the applications [Shabtai, 2009b].

Schmidt et al. (2008) focus on monitoring events at the kernel; that is, identifying critical kernel, log file, file system and network activity events, and devising efficient mechanisms to monitor them in a resource limited environment. They demonstrated their framework on static function call analysis and performed a statistical analysis on the function calls used by these applications.

---
[14] http://sourceforge.net/projects/tripwire/

Chaudhuri (2009) presented a formal language for describing Android applications and data flow among application's components. This formal language is used in [Adam, 2009] which presented the ScanDroid. ScanDroid statically analyzes Android applications and data flow between applications and compares those with security specifications defined in the application's manifest. This provides the ground for security decisions such as is the application safe and does it do what it claimed to do. ScanDroid provides the means for a developer to certify is application, and for the user to verify the proof of the certification before installation.

Thus such an approach is closely coupled with certification and can provide an automated alternative as a part of the certification process, but not only. Such a method can be used for rapid examination of Android packages and informing Google team, via the Android Market of suspicious applications.

## 8. PROPOSED MITIGATION STRATEGIES

Referring to the aforementioned 5 threat clusters, we assessed the mitigation level and effort required for applying various countermeasures for each threat cluster (Table IV).

**Threat-1: Compromising availability and/or confidentiality and/or integrity by maliciously using the permissions granted to an installed application.**

- **Intrusion Detection/Prevention System**

  An IDS solution is well-suited for defining normal behavior of the system, application or the user and detecting deviations or, alternatively, detecting malicious behavioral patterns of malware. IDS could also serve as an effective tool in discovering initially unknown and isolated threats. However, since malware can quickly adapt and mask its behavior according to the security tools capable of detecting it, the effectiveness of the IDS may decrease over time.

- **Firewall**

  The firewall is a solution for network-related attacks. It can prevent data leakage by an installed malware. However, not all attacks that abuse permissions are network-based and therefore a firewall would be very useful against a partial set of attacks.

- **Application Certification**

  Certification is an effective countermeasure against malicious applications. As each application would have to be thoroughly tested and reviewed prior to certification and permission to use any features of the device, malicious applications should be caught in their early phase and be unable to receive a proper certification. Unfortunately, nothing comes without a cost; with application certification, the costs incurred by the need to establish and maintain the certificate provider; modify the existing Android platform to support the required functionalities; and to verify each application is quite high.

- **Selective Android permissions**

  Providing the ability to approve only a subset of permissions to an installed application would reduce the risk of maliciously using granted permissions. This solution is more suitable to advanced users, and naïve users might still install applications without validating the requested permissions.

- **Automated Static Analysis and Code Verification**
  Providing the ability to automatically evaluate the nature of an application, its capabilities and the difference between what the application can do and claims to do.

**Threat-2: Compromising availability and/or confidentiality and/or integrity by an application exploiting a vulnerability in the Linux kernel or system libraries.**
- **Access Control (SELinux)**
  SELinux is well-suited for limiting the abilities of entities in the operating system. The hazardous potential of exploitable vulnerabilities could lead to a situation where the whole could be undermined if the attacker were able to obtain super-user privileges. By limiting the abilities of root processes and otherwise potentially vulnerable or high-priority entities, SELinux would prevent the attacker from forcing the system to do his/her bidding and so render the attack much less effective. However, since each entity requires the ability to execute certain commands for its normal operation, those commands must not be blocked by SELinux. In case the entity has been compromised, the attacker would still have the same maneuvering space for unleashing an attack. In other words, the attack can only be partially mitigated.

**Threat-3: Compromising availability and/or confidentiality and/or integrity of private/confidential content**
- **Login**
  The requirement of providing a secret in order to unlock certain functionalities of the device is a well-known and effective tool against a variety of threats, in particular the exposure of private content. In case the device is stolen with the lock in place, an attacker would not be able to access any of the private information without the secret. However, if the device is stolen after it is unlocked, the defense mechanism is rendered useless and the attacker could do whatever he/she wishes with the device.
- **Firewall**
  A firewall could very well protect against leakage of information through any network interface. Using either stateless or stateful inspection of content on the communication medium, it could decide whether confidential information is being sent and block the communication. As the Firewall operates in the lowest levels of the kernel, it cannot be bypassed by malicious applications (in the absence of exploitable vulnerabilities in the Linux kernel or system libraries). It can also work hand-in-hand with an access control mechanism, such as SELinux, to provide a higher level of protection. Nevertheless, network interfaces are not the only path malware can take in order to leak private data from the device; an alternative approach would be3 to send the data through SMS/MMS messages. Unfortunately, firewall cannot block such an approach.
- **Data Encryption**
  The encryption of data is an excellent means of countering exposure of private data. As only the owner knows the key that is able to decipher the data, the information would be secure in case of exposure. Even if the device was stolen and the attacker would have full access to all of the information, he/she would not be able to decypher the encryption in a reasonable time.
- **Context Aware Access Control**
  A context aware access control (CAAC) mechanism could limit access to private data depending on the context in which the device is in, based on its location, the cellular network, whether it is connected to Wi-Fi and more. Such a mechanism could defend against a variety of information disclosure attacks depending, however, on the surrounding circumstances. If the attack occurs while the device is in a context that allows access to the information, access will be permitted and the information disclosed. However, if, for example, the device is stolen and transferred to a foreign location, the data would be secure and inaccessible to the attacker.

- **Remote Management**

  Remote management capabilities are severely limited. However, when combined with additional security solutions such as a firewall or context-aware security, the security potential increases substantially. If the device is stolen, information could be protected by remotely turning on a defensive mechanism. Even during the everyday operation of the device, if the remote manager is able to identify a worm prowling the cellular / wireless network, he or she could configure the firewall accordingly to block the worm in order to prevent any information disclosure. Nevertheless, all of the above depends on human intervention during the course of an attack or prior to it. Moreover, in order to defend against attacks at the right time, a constant means of monitoring the device is required. Such a requirement is likely to be costly in terms of device resources and demanding upon the remote manager.

**Threat-4: Draining resources**

- **Resource Management**

  The resource management security solution mitigates the threat of resource drainage by malicious applications. The operation of the mechanism consists of fairly allocating resources to applications according to their needs and taking their importance into account (e.g. the phone application is very important and thus should receive more CPU than a game). In such a case, unsupervised resource drainage is not possible. If disk storage quotas are maintained and disk and network I/O is rate limited and permitted up to a certain quota, then a denial-of-service attack can be fully mitigated. However due to the difficulty of implementing such a security solution, its applicability is rather low.

- **Intrusion Detection/Prevention System**

  A host-based IDS can counter malicious drainage of battery, memory or CPU by detecting abnormal rate changes in resource levels. In practice, any malware aims to remain undetected and therefore the normal usage profile should be continuously maintained and validated.

**Threat-5: Compromising internal/protected network**

- **Virtual Private Network**

  A virtual private network (VPN) solution relies on mature principles such as message authentication codes and encryption to protect the communication.

- **Remote Management**

  The enforcement of a security policy when dealing with internal/protected networks can be done very easily by a centralized remote management framework that is controlled by the network administrator. However, the effectiveness of threat mitigation depends on the vigilance of the administrator, i.e. a human factor, which is the chink in the solution's armor.

- **Context Aware Access Control**

  When dealing with protected networks, context aware access control can in fact be viewed as an automated version of the remote management approach. Upon the detection of a context involving an active connection to the protected network, the

CAAC mechanism can increase the active security measures on the device. Such measures might include connection encryption, authentication and more.

Table IV. Mapping countermeasures to Android high-risk threats.

| | Solution | Mitigation Level | Effort |
|---|---|---|---|
| **Threat cluster-1**: Maliciously using permissions granted to an installed application | HIDS | ◐ | Medium |
| | Firewall | ◐ | Low |
| | Application certification | ● | High |
| | Selective Android permissions | ◐ | Low |
| | Automated Static Analysis and Code Verification | ◕ | Medium |
| **Threat cluster-2**: Exploiting a vulnerability in the Linux kernel or system libraries | Linux Access Control (SELinux) | ◕ | Low |
| **Threat cluster-3**: Private content | Login | ◕ | Low |
| | Firewall | ◕ | Low |
| | Data encryption | ● | Low |
| | Context Aware Access Control | ◕ | Medium |
| | Remote Management | ◕ | Medium |
| **Threat cluster-4**: Draining resource | Resource Management | ● | High |
| | HIDS | ◐ | Medium |
| **Threat cluster-6**: Compromise internal/protected network | VPN | ● | Low |
| | Remote Management | ● | Medium |
| | Context Aware Access Control | ◕ | Medium |

## 9. SUMMARY AND CONCLUSIONS

In this paper we analyzed the security issues pertaining to Google's Android and performed a systematic risk analysis in order to identify security flaws that should be mitigated in a security solution for Android devices. We highlighted potential weak points in the framework such the Android's application permission mechanism, Linux kernel, native libraries, Dalvik VM, connectivity media, and hardware. The risk arising from these vulnerabilities is amplified by the fact that as a smartphone, Android devices are expected to handle personal data and provide PC-compliant functionalities, thereby exposing the user to all the attacks that threaten users of personal computers.

We also reviewed a collection of security-related tools/mechanisms that are inherently integrated in the Android framework. Our review indicates that the defensive shell around Android was designed with extensive care since the security mechanisms embedded in Android address a broad range of security threats. Google has implemented the Portable Operating System Interface (POSIX) for preventing different applications from affecting each other. Setting each application as a different user prevents file access, signals and also distributes CPU consumption fairly with the default settings in the selected kernel. Additional security features are provided through the permission-granting mechanism that enforces restrictions on the specific operations that a particular application can perform. Signing applications is another significant security feature in which all of the application's files are signed along with their meta-data in the apk, including but not limited to the list of permissions each applications provides and requires.

Despite these Android-integrated measures, we employed a methodological qualitative risk analysis and identified five high-risk threats that should be dealt with. The main security issue, we noted, is the fact Android is an open-source platform whose source code was published after the first Android-powered devices were released to the market. This increased the chance of revealing vulnerabilities in low-level components (such as in the Linux kernel, in system's core libraries or the Dalvik VM). Moreover, several vulnerabilities were identified in the Android permission mechanism which very much increases the risk of malware installation.

Next, we surveyed additional security mechanisms that can be applied on Android-based handsets. Several of these mechanisms were tested and evaluated in our mobile device security laboratory. As a case in point, we ported SELinux into Android and activated a security policy for enhancing the protection of system processes. Moreover, a netfilter-based firewall was enabled that could be configured via a user-friendly, Android-compliant interface. We also started investigating sophisticated methods for locking Android-handsets and developing and evaluating an Intrusion Detection System based-on anomaly detection (termed Andromaly). The evaluation of the "Andromaly" IDS was performed using malicious Android applications that were developed in our laboratory especially for that purpose.

A security suite for mobile devices or smartphones (especially open-source) such as the Android should include a collection of tools, optionally operating in collaboration. Guo et al. (2004) describe defense solution space including smartphone hardening approaches (OS/hardware hardening and attack surface reduction); Internet-side defense, telecom-side defense (i.e., abnormal call blocking rate); and coordination mechanisms between Internet and a telecom infrastructure (i.e., smart-phone OSs to submit SIM IDs to Access Point (AP) for authentication when accessing the Internet).

Last, we propose several security mechanisms that can mitigate high-risk threats. We evaluated the selected countermeasures (presented in Table IV) based on the following three parameters in order to rank and prioritize them: (1) mitigation level; (2) the effort required for implementation; and (3) the number of threats that the evaluated countermeasure tackles.

It is highly important to incorporate a mechanism that can prevent or contain potential damage deriving from an attack on the Linux kernel layer such as the SELinux access control mechanism. Also, better protection should be added for hardening the Android permission mechanism or for detecting misusage of granted permissions. Consequently we assign the highest priority to SELinux along with a firewall, Intrusion Detection System, Automated Static Analysis and Code Verification and the Context Aware Access Control solutions. In a lower priority we set the Data Encryption and Selective Android Permission mechanisms. The Remote Management, VPN and Login solutions are recommended to provide a competitive edge for a Telecom operator when targeting corporate customers.

## REFERENCES

ADAM P.F. CHAUDHURI, A., AND FOSTER, J.S., 2009, SCanDroid: Automated Security Certification of Android Applications, Submitted to *IEEE Symposium of Security and Privacy*.
BOSE, A., HU, X., SHIN, K.G., AND PARK, T. 2008. Behavioral detection of malware on mobile handsets. In *Proceeding of the 6th international Conference on Mobile Systems, Applications, and Services*.

APPENDIX A – LIST OF IDENTIFIED THEATS

1. Abuse of costly services and functions (e.g., sending SMS/MMS, making phone calls, or redirecting phone calls to high rate numbers) by remotely exploiting a vulnerability in core component that is exposed on the Internet.
2. Malicious activity against a network or a network device (e.g. sending SPAM, infecting other devices, sniffing, scanning) by remotely exploiting a vulnerability in core component that is exposed on the Internet.
3. Abuse of costly services and functions (e.g., sending SMS/MMS, making phone calls, or redirecting phone calls to high rate numbers) by an application exploiting a vulnerability in core component.
4. Malicious activity against a network or a network device (e.g. sending SPAM, infecting other devices, sniffing, scanning) by an application exploiting a vulnerability in core component.
5. Abuse of costly services and functions (e.g., sending SMS/MMS, making phone calls, or redirecting phone calls to high rate numbers) by maliciously using the permissions granted by the owner at installation.
6. Malicious activity against a network or a network device (e.g. sending SPAM, infecting other devices, sniffing, scanning) by maliciously using the permissions granted by the owner at installation.
7. Disabling applications or the device by remotely exploiting a vulnerability in core component that is exposed on the Internet.
8. Disabling applications or the device by an application exploiting a vulnerability in core component.
9. Disabling applications or the device by maliciously using the permissions granted by the owner at installation.
10. Corrupting or modifying private content; or, blocking, modifying or eavesdropping on the device's communication network (e.g. phone calls, Internet communication, emails or SMS/MMS) by remotely exploiting a vulnerability in core component that is exposed on the Internet
11. Corrupting or modifying private content; or blocking, modifying or eavesdropping on the device's communication network by an application exploiting a vulnerability in core component.
12. Corrupting or modifying private content; or blocking, modifying or eavesdropping on the device's communication network by maliciously using the permissions granted by the owner at installation.
13. Obtaining, corrupting or modifying private content when browsing to a malicious web-site.
14. Blocking, modifying or eavesdropping on the device's communication network when connected to an unreliable network.
15. Receiving SPAM SMS, MMS or emails.
16. Pushing advertisements to the browser application when browsing the Internet.
17. Loss of hardware components.
18. Causing a malfunction in hardware components hardware components.

## APPENDIX B – LIST OF POTENTIAL VULNERABILITIES

V1: Android source code is free and available to everyone including hackers that can review and detect exploitable bugs and vulnerabilities.
V2: Android is an open-platform that provides API to most of the software and hardware components (SIM card, battery, memory).
V3: Integration of native system libraries and Linux kernel modules likely to be susceptible to bugs and vulnerabilities such as buffer-overflows.
V4: Android has been proven to be vulnerable to existing Linux malware (e.g. root-kits, key-loggers).
V5: Core system services and drivers run by the root-user.
V6: Unawareness of a device owner to risks of installing applications from un-trusted sources.
V7: Unawareness of a device owner to the importance of permission-granting during application installation.
V8: Unawareness of the device owner to the risk of connecting to un-trusted Wi-Fi networks and web-sites.
V9: Unawareness of the device owner to the risks posed by improper configuration (e.g., Bluetooth settings, browser settings).
V10: Unawareness of the owner to social engineering attacks.
V11: Installing applications via the Android Debug Bridge (adb-install) bypasses user interaction that exists when using regular installation methods.
V12: Impossible to grant a subset of the permissions requested by the application at installation time (all or nothing).
V13: Some of the functionalities provided by the application framework are insufficiently protected and can be maliciously used by applications.
V14: Short time-to-market of new applications may result in improperly tested and un-secured applications.
V15: Android exposes system settings that can be modified by the owner or installed applications.
V16: The browser application exposes the device to Internet-based attacks.
V17: SQLite database is commonly used in Android applications and might expose them to the SQL injection.
V18: Multiple communication transports (Bluetooth, Wi-Fi, GPRS, UMTS, Cable, PPP) provide many opportunities for malware to infiltrate a device and be activated, thus the device becomes harder to protect.
V19: Wireless communication can be eavesdropped remotely without physical access to the device.
V20: Content on the device or on the SD card is not encrypted.
V21: Android is not equipped with proper attack countermeasure mechanisms (such as encryption of data, firewall, identification and authentication).
V22: Android devices can be used to attack other devices, computers or networks (e.g. by running scanners, SMS/MMS/email worms, inject Trojans via USB cable).
V23: There is no disk storage quota per application.
V24: There is no memory (RAM) quota per application.
V25: Android device, like any other mobile phone, is small in size and thus more prone to theft and loss.
V26: The device is vulnerable to environmental factors such as humidity and heat.